\documentclass{jpp}
\usepackage{graphicx}
\usepackage{subcaption}
\usepackage[utf8]{inputenc}
\usepackage[T1]{fontenc}
\usepackage{amsmath}

\usepackage[colorlinks=true, allcolors=blue]{hyperref}
\hypersetup{
    colorlinks=true,       
    linkcolor=cyan,          
    citecolor=magenta,        
    filecolor=magenta,      
    urlcolor=cyan,           
    runcolor=cyan
}

\shorttitle{Simulating nonlinear optical processes}
\shortauthor{Y. Shi, B. Evert and others}

\title{Simulating nonlinear optical processes on a superconducting quantum device}

\author{Yuan Shi\aff{1}
  \corresp{\email{Yuan.Shi@colorado.edu}},
  Bram Evert\aff{2},
  Amy F. Brown\aff{3},
  Vinay Tripathi\aff{3},
  Eyob A. Sete\aff{2}, 
  Vasily Geyko\aff{4},   
  Yujin Cho\aff{4},   
  Jonathan L DuBois\aff{4}, 
  Daniel Lidar\aff{3,5}, 
  Ilon Joseph\aff{4}, 
  \and Matt Reagor\aff{2}
  }

\affiliation{
\aff{1}Department of Physics, Center for Integrated Plasma Studies, University of Colorado Boulder, Boulder, CO 80309, USA
\aff{2}Rigetti Computing, 775 Heinz Ave, Berkeley, California, 94710 USA
\aff{3}Department of Physics \& Astronomy and Center for Quantum Information Science \& Technology, University of Southern California, Los Angeles, California 90089, USA
\aff{4}Lawrence Livermore National Laboratory, Livermore, California 94550, USA
\aff{5}Department of Electrical \& Computer Engineering and Department of Chemistry, University of Southern California, Los Angeles, California 90089, USA
}

\begin{document}

\maketitle

\vspace{-5pt}
\begin{abstract}
Simulating plasma physics on quantum computers is difficult because most problems of interest are nonlinear, but quantum computers are not naturally suitable for nonlinear operations. In weakly nonlinear regimes, plasma problems can be modeled as wave-wave interactions. In this paper, we develop a quantization approach to convert nonlinear wave-wave interaction problems to Hamiltonian simulation problems. We demonstrate our approach using two qubits on a superconducting device. 
Unlike a photonic device, a superconducting device does not naturally have the desired interactions in its native Hamiltonian. Nevertheless, Hamiltonian simulations can still be performed by decomposing required unitary operations into native gates. 
To improve experimental results, we employ a range of error mitigation techniques. Apart from readout error mitigation, we use randomized compilation to transform undiagnosed coherent errors into well-behaved stochastic Pauli channels. Moreover, to compensate for stochastic noise, we rescale exponentially decaying probability amplitudes using rates measured from cycle benchmarking.   
We carefully consider how different choices of product-formula algorithms affect the overall error and show how a trade-off can be made to best utilize limited quantum resources. This study provides an example of how plasma problems may be solved on near-term quantum computing platforms.
\end{abstract}

\section{Introduction\label{sec:Intro}}
The physics of plasmas is often multi-scale, multi-physics, and highly nonlinear. While classical computers have no trouble handling nonlinearity, the multi-scale and multi-physics aspects make many plasma problems challenging even for classical supercomputers. Simulations of larger systems with finer resolutions are restricted by both the memory and time constraints of classical computers.
In comparison, quantum computers, which are still under development, in principle have an exponentially larger memory, and thus have attracted substantial interest in recent years, including within the plasma community \citep{dodin2021applications,joseph2023quantum,koukoutsis2023quantum}.
In addition to a larger memory, quantum computers support algorithms that offer quadratic to exponential speedups over the best-known classical algorithms for specialized problems, such as quantum search, Fourier transform, and factoring. However, whether quantum computers have the potential to offer a speedup for plasma problems remains an open question.
A major difficulty is that quantum algorithms typically rely on unitary operations, and quantum computers are not naturally suited for general nonlinear operations needed in plasma physics. Attempts have been made to develop schemes that can handle nonlinear plasma problems on quantum computers \citep{joseph2020koopman,engel2021linear,liu2021efficient,lin2022koopman}. Many of these schemes rely on future fault-tolerant quantum computers and thus cannot be tested on current noisy quantum devices. While developing abstract algorithms for future quantum hardware is valuable, performing concrete examples on current devices allows the community to build intuition about how quantum computation works in practice and reveals potential gaps between theoretical expectations and the reality of present-day capabilities.

In this paper, we show that small initial value problems that model the evolution of nonlinear wave-wave interactions can be solved on current quantum hardware, after employing a suite of error mitigation techniques. Wave-wave interactions are a general framework for weakly nonlinear dynamical systems \citep{davidson2012methods, zakharov1992kolmogorov,nazarenko2011waveturbulence}. Many dynamical systems possess fixed points, which correspond to equilibrium states of the system. When slightly perturbed away from a stable fixed point, the system oscillates. Small amplitude oscillations of fields are usually called linear waves in plasma physics. At the linear level, the waves are decoupled eigenmodes. However, at larger amplitudes, waves become coupled due to nonlinearities of the system, leading to what are known as wave-wave interactions.  
For example, laser-plasma interactions, such as Raman and Brillouin scattering, are often described as three-wave interactions \citep{michel2023introduction}, because they involve an incoming laser, an outgoing laser, and a mediating plasma wave.
As another example, filamentation and modulational instabilities of plasma waves can be described as four-wave interactions \citep{michel2023introduction}. In these instabilities, fluctuations of the laser amplitude depend on the laser intensity. The instability is a four-wave process because the laser beats with its fluctuations to produce side bands. 
Wave-wave interactions arise not only in laser-plasma interactions, but also occur in magnetically confined plasmas \citep{hansen2017parametric, nazarenko2011waveturbulence}, astrophysical plasmas \citep{bowen2018density}, and many other physical systems \citep{zakharov1992kolmogorov, nazarenko2011waveturbulence}.

While plasma physics often treats wave-wave interactions classically, we note that these nonlinear processes are intrinsically quantum. In fact, Raman \citep{raman1928negative} and Brillouin \citep{brillouin1914light} scattering were first studied in gases and solids as quantum processes before these terms were borrowed by plasma physicists. For example, at the quantum level, a single incoming photon can scatter into a single outgoing photon by emitting a single phonon \citep{bowen2015quantum}. When there are many indistinguishable photons, the scattering probability amplitudes add up to determine the total scattering probability, but the underlying three-wave interaction is not fundamentally different. In the study of laser-plasma interactions, laser light is often treated as classical electromagnetic waves. However, it is worth noting that from a quantum perspective, laser light is just a special collection of photons known as a {\it coherent state}, for which photon number distribution is Poissonian. In this regard, laser scattering from plasma is just a special example of three-wave interactions. 
More generally, photons can occupy other quantum states \citep{loudon2000quantum}. For example, other important classes of quantum states are known as  {\it squeezed states} \citep{breitenbach1997measurement}, which are different from coherent states because their probability distribution functions are not Poissonian. 
While laser light may be approximated as classical electromagnetic waves, squeezed light cannot be described by classical waveforms at all. If one attempts to assign a waveform to a squeezed state, then the waveform would have to fluctuate from cycle to cycle, and yet its Fourier transform must still have a narrow peak.
To model quantum light interacting with matter, one can no longer use a classical treatment of wave-wave interactions. To capture quantum interference, which results from summing probability amplitudes rather than the probabilities themselves, one must use a quantum treatment.

The high cost associated with modeling the exact quantum-mechanical evolution makes it desirable to develop quantum simulation capabilities for wave-wave interactions induced by quantum light. At the same time, the intrinsic quantum nature of wave-wave interactions makes it natural
to study these nonlinear processes using a quantization approach \citep{shi2017three,shi2021simulating}. 
After promoting classical amplitudes to quantum creation and annihilation operators, the nonlinear equations that are used to describe wave-wave interactions are derivable as Heisenberg equations. By choosing a convenient basis, the Hamiltonian operators can be converted to finite-dimensional Hamiltonian matrices, and we perform quantum Hamiltonian simulations in the Schr\"{o}dinger picture. The goal of the simulations is to extract observables at some later time, which can be constructed using the quantum state we evolve. 
In the limit of large occupation number, the quantized model reproduces linear instabilities in classical plasmas \citep{May23}. 

For quantum Hamiltonian simulations to be efficient, a number of criteria must be met. One criterion is that the Hamiltonian matrices must have special properties. In our case, the matrices are sparse, for which efficient quantum algorithms exist \citep{berry2007efficient,berry2014exponential,low2017optimal}, at least for future error-corrected quantum computers. Another criterion is that one must not be interested in the full information stored in the quantum memory. In our cases, we are interested, for example, in the intensity of the backscattered light, which involves just a few expectation values. 
Overall, the scheme we develop is efficient for simulating how quantum light interacts with plasmas and can accelerate such simulations on future error-corrected quantum computers.

We demonstrate our scheme on a current superconducting quantum device using a two-qubit example. Our demonstration pushes the limit of what current hardware can do and addresses important trade-off between different error sources for product-formula algorithms. Rather than implementing black-box quantum Hamiltonian simulation algorithms, which are not feasible on current devices and will only become more efficient for larger system sizes, we simply decompose the requisite unitary operations into elementary gates using a Cartan decomposition \citep{smith2020open, huang_quantum_2023}, which is the most efficient approach for two-qubit problems. For more qubits, the number of two-qubit gates Cartan decomposition requires grows exponentially, so implementing larger problems on future devices require different methods. 
In our previous work, we have demonstrated that current devices can perform two-qubit time evolution using the exact unitary \citep{shi2021simulating}, which is obtained by analytically exponentiating the full Hamiltonian matrix. In principle, if one could perform the exact exponentiation, then the problem is already solved. However, even when the exact unitary was prescribed, we found that quantum devices had difficulties repeating the dense unitary beyond a few cycles, unless one can drastically improve gate fidelity.

In this paper, we demonstrate the next level of quantum computation by assuming that only certain parts of the exact unitary can be performed efficiently. More specifically, since our Hamiltonian has two non-commuting terms, we assume that each term can be exponentiated exactly. Then, we use product formulas to approximate the total unitary. Product formulas are often known as Hamiltonian splitting in the plasma literature \citep{he2015hamiltonian,morrison2017structure}, and are known as Lie-Trotter-Suzuki decomposition in the quantum literature \citep{trotter1959product,suzuki1976generalized}. On future quantum computers, quantum Hamiltonian simulations may still utilize product formulas, when the full Hamiltonian cannot be simulated efficiently but its subparts can.

For current devices, which do not yet support operational error correction, we show that reasonable results can be obtained only after we employ a range of error mitigation and suppression techniques.
First, we suppress the occurrence of errors by addressing the highest performing two-qubit gates on the device. Rather than using an off-the-shelf gate, we calibrate a $\sqrt{\text{iSWAP}}$ (SQISW) gate \citep{abrams_implementation_2020} which provides superior performance and expressiveness \citep{peterson_fixed-depth_2020} for our problem.
Second, we mitigate readout errors, which occur if the qubit states are misclassified during measurements. We experimentally measure the confusion matrix, which describes the probability of preparing a qubit register in one state but classifying it in another state. We estimate the true distribution of samples from a noisy one using iterative Bayesian unfolding \citep{nachman2020unfolding}. 
Third, we mitigate coherent gate errors, which occur if the realized unitary gate systematically differs from the target unitary. Coherent errors often arise from drifts in system parameters, so that gates calibrated at one time no longer remain perfectly calibrated at a later time. Coherent errors are problematic because they accumulate at each gate operation and can interfere constructively. 
To mitigate coherent errors, we use randomized compilation \citep{Wallman16, hashim2020randomized}. When a unitary operation is called, the hardware implements it using a different but equivalent gate sequence, sampled at random from a precompiled set of choices. This technique converts coherent errors into stochastic errors, thereby suppressing constructive interference between errors.
Finally, to mitigate incoherent errors, which occur both due to the random selection of gate sequences and intrinsic quantum decoherence on the hardware, we employ an amplitude rescaling technique \citep{Ville22}. The technique assumes that the probability of a pure quantum state decays exponentially to the fully mixed state. We measure the decay rate using cycle benchmarking \citep{erhard2019characterizing}, and compensate for the decay by multiplying the probabilities with an exponential growth factor. While this technique partially recovers signals, it also leads to an exponential growth of error bars. The maximum simulation time and the maximum achievable gate depth is reached when the error bars become comparable to the signal of interest.

Using error mitigation techniques, we substantially extend the achievable simulation depth, which is nevertheless still limited. 
After error mitigation, we can accurately measure observables at a depth of about two hundred two-qubit gates, for a reasonable shot budget. Two-qubit gates, which require longer hardware runtime and are more vulnerable to decoherence, have significantly lower fidelity than single-qubit gates, and are the limiting factor on current quantum devices. 
Given that errors grow with the gate depth, more accurate results may be obtained for a targeted final time by reducing the time step size, which in turn reduces the discretization error per step at the expense of increasing the accumulation of hardware errors. 
Alternatively, more accurate results may be obtained by using a higher order algorithm, which decreases the algorithmic error per step at the expense of more gates per step, but allows the use of larger time step sizes. 
We carefully consider different choices and show that a trade-off can be made to best utilize limited quantum resources.

The paper is organized as follows. In Sec.~\ref{sec:model}, we discuss classical models of laser-plasma interactions and show how the model can be quantized and converted to a Hamiltonian simulation problem for quantum computers. In Sec.~\ref{sec:exact}, we implement an exact two-qubit problem on superconducting hardware, and show that error mitigation techniques meaningfully improve simulation results. In Sec.~\ref{sec:trotter}, we investigate product formula approximations and discuss how to best utilize limited quantum resources.

\section{Classical and quantum models of laser pulse compression\label{sec:model}}
An important class of wave-wave interactions in plasma physics are laser-plasma interactions (LPIs). In this paper, we consider an example scenario where a plasma is used for laser pulse compression \citep{malkin1999fast}, during which the intensity of a seed laser pulse is amplified while its duration is shortened. Other LPI scenarios, as well as other cases of wave-wave interactions, may be treated in a similar fashion. We will first describe how the problem is usually treated classically in plasma physics, and then develop a quantized model that is amenable to simulations on quantum computers.

\subsection{Classical model}
Laser amplification is often treated as a parametric process, where the signal and idler waves grow by consuming a pump wave. When the pump energy is being replenished, or when the pump energy dominates, one may approximate the pump amplitude $a_1$ as a constant, in which case the seed amplitude $a_2$ and the idler amplitude $a_3$ grow exponentially. However, when the pump amplitude is not held constant, the three-wave nature of the underlying interaction becomes apparent. The interaction is often described by the three-wave equations \citep{davidson2012methods}
\begin{subeqnarray}
    \label{eq:three-wave}
    d_t a_1 &=& g a_2 a_3, \\
    d_t a_2 &=& -g^* a_1 a_3^\dagger, \\
    d_t a_3 &=& -g^* a_1 a_2^\dagger,
\end{subeqnarray}
where $d_t$ is the advective derivative, $g$ is the coupling coefficient, $g^*$ is its complex conjugate, and $a^\dagger$ denotes the complex conjugate of $a$ in the classical model. 
The advective derivative is specific for each wave and is defined as $d_t=\partial_t + \mathbf{v}_g\cdot\nabla + \mu$, where $\mathbf{v}_g=\partial\omega/\partial\mathbf{k}$ is the group velocity and $\mu$ is the damping rate of the wave.  
The complex-valued amplitude $a$ is the slowly varying envelope of the classical wave. The amplitude is normalized such that $n=|a|^2$ is proportional to the wave action density, which is proportional to the number of photons in the wave.

Three-wave interactions satisfy a number of conservation laws. First, the equations describe resonant interactions where both energy and momentum are conserved. For waves with narrow bandwidth, their 4-momentum density is proportional to $k^\mu=(\frac{\omega}{c},\mathbf{k})$. Energy and momentum are conserved because the interaction satisfies the resonance conditions $k_1^\mu=k_2^\mu+k_3^\mu$. 
Moreover, the interaction has two independently conserved actions $S_2=n_1+n_2$ and $S_3=n_1+n_3$. These actions are constants of motion because their advective derivatives are zero. The physical meaning of action conservation is that the three-wave interaction splits each pump photon into a seed and an idler photon. So, whenever $n_1$ decreases by one, both $n_2$ and $n_3$ increase by one, and vice versa.

The three-wave equations are partial differential equations that describe how the wave amplitudes evolve in both space and time. If all amplitudes are initially uniform in space, then they will remain uniform as they evolve in time. However, if the amplitudes are not uniform, two competing effects change the envelopes of the waves. 
First, wave advection transports the wave envelope in the direction of the group velocity. The advection is a linear process, and the envelope remains unchanged in the co-moving frame. 
Second, the three-wave interaction changes amplitudes locally. Since the interaction is nonlinear, the change is faster where the amplitudes of the other two waves are larger.

Laser pulse compression is a special scenario where the competition between the two effects lead to the amplification and shortening of the seed wave. 
At later stages of pulse compression, the intensity of the seed far exceeds the pump. The large $a_2$ induces an additional relativistic nonlinearity. The nonlinearity originates from the fact that, in plasmas,
photons are massive particles due to their interactions with free charges. In unmagnetized plasmas, photons satisfy the dispersion relation $\omega^2=\omega_p^2+c^2k^2$, where the photon mass can be identified with the plasma frequency $\omega_p=(e^2n_e/\epsilon_0 m_e)^{1/2}$. 
Here, $e$ is the electron charge and $n_e$ is the electron density. 
Because electrons oscillate in the laser's electric field, the effective electron mass $m_e$ is replaced by $\gamma m_e$ when the electron quiver speed $v_q$ becomes comparable to the speed of light $c$, where $\gamma=1/\sqrt{1-v_q^2/c^2}$.
As the seed pulse propagates, at places where the pulse is more intense, the photon mass $\omega_p \propto \gamma^{-1/2}$ becomes smaller. A smaller $\omega_p$ leads to a larger $k$ at a fixed $\omega$, which means a larger group velocity $\mathbf{v}_g=c^2\mathbf{k}/\omega$. Consequently, the more intense part of $a_2$ moves at a higher group velocity. If the envelope of $a_2$ has initial modulations, then they will pile up and grow. This process is known as relativistic modulational instability.

The modulational instability may be understood as a four-wave process, where the laser beats with its modulations to produce side bands. In the weakly relativistic limit, we can expand $1/\gamma\simeq  1-v_q^2/(2c^2)$. Since electrons respond primarily to the electric field of the laser pulse, Newton's equation $m_e d\mathbf{v}_q/dt\simeq e\mathbf{E}$ becomes $\tilde{\mathbf{v}}_q\simeq \textrm{i}e\tilde{\mathbf{E}}/(m_e\omega)$ in Fourier space. Denoting the normalized laser amplitude by $\boldsymbol{\alpha}=e\tilde{\mathbf{E}}/(m_e\omega c)$, then the average quiver speed is $\langle v_q^2/c^2\rangle=\frac{1}{2} e^2|\tilde{E}|^2/(m_e\omega c)^2= \frac{1}{2}|\alpha|^2$. Replacing $\omega_p^2\rightarrow \omega_p^2/\gamma \simeq \omega_p^2(1-\frac{1}{4}|\alpha|^2)$, the photon dispersion relation is approximated.
The dispersion relation is derivable from the wave equation $[\partial_t^2-c^2\nabla^2+\omega_p^2(1-\frac{1}{4}|\alpha|^2)]\mathbf{E}=0$.
In the WKB approximation, the complex wave is $\mathbf{E}(x,t)=\mathbf{A}(x,t)e^{\textrm{i}\theta}$, where $\theta=\mathbf{k}\cdot\mathbf{x}-\omega t$, and the envelope $\mathbf{A}$ varies slowly in the sense that $|\partial_t\mathbf{A}/\mathbf{A}|\ll\omega$ and $|\nabla \mathbf{A}/\mathbf{A}|\ll k$.
Then, to leading order, the wave equation is approximated by $[\partial_t + \mathbf{v}_g\cdot\nabla -\textrm{i}\omega_p^2|\alpha|^2/(8\omega)] \mathbf{A}=0$.
When focusing on the scalar amplitude $A$, because $A\propto\alpha\propto a$, the equation that describes the modulational instability of the seed pulse is 
\begin{equation}
    \label{eq:modulation}
    d_t a_2 = \textrm{i} R a_2^\dagger a_2 a_2,
\end{equation}
where $d_t$ is again the advective derivative and $R=\omega_p^2/(8\omega$) is the coupling coefficient. For relativistic modulational instability, $R>0$ is a real number, which means $iR |a_2|^2$ is purely imaginary. The above equation thus modulates the phase of the complex $a_2$ in such a way that a larger $|a_2|$ leads to a faster phase evolution.

Laser pulse compression is described by the combined Eqs.~(\ref{eq:three-wave}) and (\ref{eq:modulation}). The modulational instability only affects $a_2$ because during laser pulse compression, the peak values satisfy $|a_2|\gg|a_1|, |a_3|$. 
Since the three-wave interaction is a phase sensitive process, the modulational instability spoils the amplification process by introducing a phase mismatch.

\subsection{Quantum model}
In the classical model, the amplitude $a$ is a complex-valued function, and $n=|a|^2$ is proportional to the number of photons. This setup naturally admits canonical quantization for bosonic quantum fields $[a_i(\mathbf{x}), a_j^\dagger(\mathbf{y})] = \delta_{ij}(2\pi)^3\delta^{(3)}(\mathbf{x}-\mathbf{y})$, where the indices $i, j=1,2,3$, and the operators have spatial dependencies. 
Since we will later implement the model on quantum hardware, which has a limited number of qubits, in this paper we will focus on the temporal problem with no spatial dependence. 
In this case, when damping is negligible, the advective derivative $d_t\rightarrow\partial_t$ is reduced to a partial derivative in time, and the operators satisfy the canonical quantization conditions
\begin{equation}
    [a_i, a_j^\dagger] = \delta_{ij}.
\end{equation}
The quantization promotes normalized amplitudes to creation and annihilation operators, and the Kronecker delta distinguishes the three types of waves in the system. For each wave type, the number operator is $n_i=a_i^\dagger a_i$. 
The eigenstates of $n_i$ are the Fock states $|m_i\rangle$, namely, $n_i|m_i\rangle = m_i|m_i\rangle$, where $m_i=0,1,2\dots$, and $|m_i\rangle=(a_i^\dagger)^{m_i}|0_i\rangle/\sqrt{m_i!}$. Here, $|0_i\rangle$ is the ground state of wave $i$, which is annihilated by $a_i|0_i\rangle=0$.
More generally, these quantum harmonic oscillators have the usual matrix elements $a_i|m_i\rangle=\sqrt{m_i}|m_i-1\rangle$ and $a_i^\dagger|m_i\rangle=\sqrt{m_i+1}|m_i+1\rangle$.
Since we have three types of waves, it is convenient to abbreviate the tensor-product state $|m_1\rangle\otimes|m_2\rangle\otimes|m_3\rangle$ as $|m_1, m_2, m_3\rangle$. This number basis is natural for the quantized problem.

While the Schr\"{o}dinger picture $\textrm{i}\partial_t|\psi\rangle=H |\psi\rangle$ is more convenient for quantum simulations, the connection between the quantum and classical models is more transparent in the Heisenberg picture $d_ta=\textrm{i}[H, a]$. 
For three-wave interactions, Eqs.~(\ref{eq:three-wave}) are the Heisenberg equations from the cubic Hamiltonian
\begin{equation}
    \label{eq:HT}
    H_T = \textrm{i}g a_1^\dagger a_2 a_3 - \textrm{i}g^* a_1 a_2^\dagger a_3^\dagger.
\end{equation}
A degenerate form of $H_T$, where $a_1=a_2$, commonly arises in optomechanical systems \citep{aspelmeyer2008focus, kong2018two, lake2020two, shang2023coupling}.
In our case, $a_1\ne a_2$. The first term of $H_T$ annihilates a seed and an idler photon to create a pump photon, while the second term of $H_T$ is the reverse process where a pump photon decays into a seed and an idler photon. 
Although the Heisenberg equations for $a_j$ are formally identical to the classical three-wave equations, the difference between the quantum and classical systems become apparent when one calculates higher order cumulants. For example, because $a_j$ and $a^\dagger_j$ do not commute, the Heisenberg equation for $n_i$ is different from its classical counterpart \citep{shi2021simulating}.
Similarly, for the four-wave interaction, Eq.~(\ref{eq:modulation}) is the Heisenberg equation from the quartic Hamiltonian
\begin{equation}
    \label{eq:HF}
    H_F = -\frac{R}{2} a_2^\dagger a_2^\dagger a_2 a_2,
\end{equation}
which is also known as the self-Kerr nonlinearity in the quantum literature \citep{kerr1875xl}. Since $R>0$ for the modulational instability, the negative sign in $H_F$ means that photons tend to condense together, which leads to a lower energy of the system. In Eq.~(\ref{eq:HF}), the operators are normal ordered $a_2^\dagger a_2^\dagger a_2 a_2=n_2^2-n_2$, which is different from other orderings such as $a_2^\dagger a_2 a_2^\dagger a_2=n_2^2$. Different orderings differ by factors of the number operator $n_2$. 
For our purposes, the ordering does not matter, because the total Hamiltonian also includes the energy of the non-interacting harmonic oscillators $H_0=\sum_j\omega_ja_j^\dagger a_j=\sum_j\omega_jn_j$. In the interaction picture of $H_0$, these quadratic terms are trivially removed, and Eqs.~(\ref{eq:HT}) and (\ref{eq:HF}) should be understood as three- and four-wave interactions in the interaction picture of $H_0$. Then, different orderings in Eq.~(\ref{eq:HF}) are equivalent up to a re-definition of $\omega_2$. Hence, it is sufficient to consider normal-ordered operators.

To use quantum Hamiltonian simulations to solve the quantized wave-wave interaction problems, we will focus on the Schr\"{o}dinger picture and use a basis that respects action conservation.
In classical wave-wave interactions, $S_2=n_1+n_2$ and $S_3=n_1+n_3$ are known as the conserved wave actions.
In the quantized model, $[H_T, S_2]=[H_T, S_3]=0$, and $H_F$ also commutes with $S_2$ and $S_3$. 
Therefore, it is convenient to use eigenstates of $S_2$ and $S_3$ as the computational basis, which we shall call the action basis.
For the laser pulse compression problem, since we are primarily interested in the seed wave $a_2$, we label the action basis by
\begin{equation}
    \label{eq:basis}
    |\phi_j^{s_2, s_3}\rangle = |s_2 - j,\, j,\, s_3-s_2+j\rangle,
\end{equation}
where $j$ is the number of photons in the seed wave and is bounded within the range $j_{\min} \le j\le s_2$ with $j_{\min}=\max(0, s_2-s_3)$. In other words, the $|\phi_j^{s_2, s_3}\rangle$ basis spans a $D=\min(s_2, s_3)+1$ dimensional subspace of $|m_1, m_2, m_3\rangle$, where $m_1=s_2-j$, $m_2=j$, and $m_3=s_3-s_2+j$. The bounds for $j$ come from the fact that $m_i\ge 0$.
The nonnegative integers $s_2$ and $s_3$ are eigenvalues of $S_2$ and $S_3$, namely, $S_2|\phi_j^{s_2, s_3}\rangle=s_2 |\phi_j^{s_2, s_3}\rangle$ and $S_3|\phi_j^{s_2, s_3}\rangle=s_3 |\phi_j^{s_2, s_3}\rangle$. In other words, the infinite dimensional Hilbert space can be decomposed as a direct sum of finite dimensional subspaces, where each subspace is labeled by a pair of quantum numbers $(s_2, s_3)$.

In the action basis, the Hamiltonian becomes block diagonal. 
The total Hamiltonian that governs the mixed three- and four-wave interaction problem is 
\begin{equation}
    H = H_T +H_F,
\end{equation}
where $[H_T, H_F]\ne 0$.
The matrix elements of $H_T$ are $H_T |\phi_j^{s_2, s_3}\rangle = \textrm{i}g \eta_{j-\frac{1}{2}}^{s_2, s_3} |\phi_{j-1}^{s_2, s_3}\rangle 
- \textrm{i}g^* \eta_{j+\frac{1}{2}}^{s_2, s_3} |\phi_{j+1}^{s_2, s_3}\rangle$, where the reduced matrix element $\eta_{j-\frac{1}{2}}^{s_2, s_3}=\sqrt{(s_2+1-j)j(s_3-s_2+j)}$ is inherited from the creation and annihilation operators. 
Notice that $H_T$ only couples nearest neighbors in the action basis. Moreover, since the annihilation operator terminates at $m=0$, the reduced matrix element vanishes both for bottom and top values of $j$.
The bottom values of $j$ are either $j=0$ or $j=s_2-s_3$, and in both cases $\eta_{-\frac{1}{2}}^{s_2, s_3}=\eta_{s_2-s_3-\frac{1}{2}}^{s_2, s_3}=0$.
The top value is $j=s_2$, for which $\eta_{s_2+\frac{1}{2}}^{s_2, s_3}=0$.
In other words, in the $|\phi_j^{s_2, s_3}\rangle$ basis, the matrix of $H_T$ is block tridiagonal, where each block is finite dimensional, with no need for artificial truncation. Within each block, $H_F |\phi_j^{s_2, s_3}\rangle = -R\zeta_j |\phi_j^{s_2, s_3}\rangle$.
The reduced matrix element is $\zeta_j=\frac{1}{2}j(j-1)$, which is nonzero only when $j\ge 2$. This is intuitive because the self-Kerr effect is a photon-photon nonlinearity, so at least two photons are needed to see the effect. 
Because the matrix for $H_F$ is diagonal, one can easily exponentiate the matrix analytically. Efficient quantum simulation algorithms also exist for diagonal Hamiltonian matrices with a smooth structure \citep{Welch_2014}.

Because $H$ is block diagonal in the action basis, we can perform Hamiltonian simulations separately in each $(s_2, s_3)$ subspace. 
We span the wave function by 
\begin{equation}
    \label{eq:expansion_general}
    |\phi(t)\rangle = \sum_{s_2, s_3=0}^{\infty} \sqrt{p^{s_2, s_3}}\sum_{j=j_{\min}}^{s_2} c^{s_2, s_3}_j(t) |\phi^{s_2, s_3}_j\rangle.
\end{equation}
The time-independent $p^{s_2, s_3}\ge 0$ is the probability that the state is within the $(s_2, s_3)$ subspace, and the total probability $\sum_{s_2, s_3}p^{s_2, s_3} = 1$. 
Because different subspaces decouple, the wave function is normalized separately within each subspace $\sum_{j} |c^{s_2, s_3}_j|^2=1$.
After performing Hamiltonian simulations for the subspaces, which can be done in parallel, one can sum their contributions using weights that are predetermined by initial conditions.

For plasma problems, the initial states are often tensor-product states $|\phi(t=0)\rangle=|\psi_1\rangle\otimes |\psi_2\rangle\otimes |\psi_3\rangle$, where the three waves are initially unentangled. For each wave, $|\psi_i\rangle=\sum_m \beta^i_m|m\rangle$ can be expanded in its Fock basis, where $p^i_m=|\beta^i_m|^2$ is the probability that the $i$-th wave occupies its $m$-th state. 
For example, for classical laser pulse compression, the pump and seed lasers are initially in a coherent state $|\xi\rangle_c$, where $\xi=r e^{\mathrm{i}\theta}$ is a complex number. For the coherent state, the probability amplitudes are $\beta_m=\exp(-\frac{1}{2}r^2) \xi^m/\sqrt{m!}$. In this case, $p_m$ is a Poisson distribution, with $\langle n \rangle = r^2$ and $\langle n^2\rangle = \langle n \rangle^2 + \langle n \rangle$, and $p_m$ peaks near $m\sim r^2$. When $m\gg 1$, $p_m\simeq \exp(-r^2) (er^2/m)^m/\sqrt{2\pi m}$ decays faster than exponential. 
As another example, for interactions between quantum light, the initial state may be a squeezed vacuum state $|\xi\rangle_q$, for which $\beta_{2m+1}=0$ for odd states and $\beta_{2m}=(\textrm{sech}\, r)^{1/2} (-\frac{1}{2} e^{\mathrm{i}\theta} \textrm{tanh}\, r)^m \sqrt{(2m)!}/m!$ for even states.
In this case, $\langle n \rangle = \textrm{sinh}^2 r$ and $\langle n^2\rangle = 3\langle n \rangle^2 + 2\langle n \rangle$. The probability $p_{2m}$ monotonically decreases, and $p_{2m}\simeq \textrm{sech}\, r (\textrm{tanh}\, r)^{2m}/\sqrt{\pi m}$ decays much slower than a coherent state but faster than an exponential. 
In both examples, due to the superexponential decays of $p_m$, it is sufficient to keep track of a finite number of states in the Fock basis. 
Expanding the initial tensor-product state in the action basis gives
\begin{equation}
    \label{eq:initial_conditions}
    \sqrt{p^{s_2, s_3}}\,c^{s_2, s_3}_j(t=0) = \beta^1_{s_2-j}\, \beta^2_j \,\beta^3_{s_3-s_2+j}.
\end{equation}
The simplest case is when only the pump wave is initially excited, which means that the seed and idler waves are initially in the ground state $\beta^2_m=\beta^3_m=\delta_{m,0}$. In this case, $p^{s_2, s_3}=|\beta^1_{s_2}|^2\delta_{s_2, s_3}$ and $c^{s_2, s_3}_j(t=0) = \delta_{j,0}$. At later time, the wave function is spanned by
$|\phi(t)\rangle = \sum_{s=0}^\infty \beta^1_s\sum_{j=0}^s c^{s,s}_j(t)|\phi^{s,s}_j\rangle$. 
Another special case is when the idler is initially in the ground state, namely, $\beta^3_m=\delta_{m,0}$. In this case, $p^{s_2, s_3}=|\beta^1_{s_3}|^2 |\beta^2_{s_2-s_3}|^2\Theta_{s_2,s_3}$, where the step function $\Theta_{i,j}=0$ when $i<j$ and $\Theta_{i,j}=1$ when $i\ge j$. The initial condition is $c^{s_2, s_3}_j(t=0)=\delta_{j, s_2-s_3}$, and at a later time the wave function is spanned by $|\phi(t)\rangle = \sum_{s_2=0}^\infty \sum_{s_3=0}^{s_2} \beta^1_{s_3} \beta^2_{s_2-s_3} \sum_{j=s_2-s_3}^{s_2} c^{s_2,s_3}_j(t)|\phi^{s_2,s_3}_j\rangle$. 
In both examples, the wave function initially occupies the ground state, namely, the $j=j_{\min}$ state, within each $(s_2, s_3)$ subspace.

For plasma simulations, the observables of interest are often the wave amplitudes, namely, the expectation values $\langle n_i \rangle$ for the three waves. 
Due to action conservation, the three amplitudes are not independent. For given initial conditions, the conserved actions are $\langle S_2 \rangle = \sum_{s_2, s_3 = 0}^\infty p^{s_2, s_3} s_2$ and $\langle S_3 \rangle = \sum_{s_2, s_3 = 0}^\infty p^{s_2, s_3} s_3$.
Because the probabilities $p^{s_2, s_3}$ are time independent, the expectation values $\langle S_2 \rangle$ and $\langle S_3 \rangle$ are constants of motion. 
The amplitudes of the pump and idler waves are related to the amplitude of the seed wave by
\begin{subeqnarray}
    \label{eq:n1n3_expectation}
    \langle n_1(t) \rangle &=& \langle S_2 \rangle - \langle n_2(t)\rangle \\
    \langle n_3(t) \rangle &=& \langle S_3 \rangle - \langle S_2 \rangle + \langle n_2(t) \rangle.
\end{subeqnarray}
To compute the seed amplitude, we use the orthonormality of the action basis, which gives $\langle n_2 \rangle = \langle\phi|n_2|\phi\rangle = \langle\phi| \sum_{s_2, s_3, j} \sqrt{p^{s_2, s_3}} c^{s_2, s_3}_j j |\phi^{s_2, s_3}_j\rangle = \sum_{s_2, s_3, j} p^{s_2, s_3} |c^{s_2, s_3}_j|^2 j$. The amplitude can be computed in two steps
\begin{subeqnarray}
    \label{eq:n1n3_expectation}
    \langle n_2^{s_2, s_3} (t) \rangle &=& \sum_{j=j_{\min}}^{s_2} j |c^{s_2, s_3}_j(t)|^2, \\
    \langle n_2(t) \rangle &=& \sum_{s_2, s_3 = 0}^\infty p^{s_2, s_3} \langle n_2^{s_2, s_3} (t) \rangle,
\end{subeqnarray}
where the first step computes the expectation value in each $(s_2, s_3)$ subspace, and the second step performs weighted sums using predetermined probabilities. 
For interactions involving quantum light, one may also be interested in observing higher-order cumulants, which can be measured in similar ways.

\subsection{Exact dynamics}
In the remaining part of this paper, we will focus on a single subspace and drop the $s_2$ and $s_3$ superscripts to keep notations compact.
When $s_2=s_3=0$, the subspace is trivial because $D=1$ is one dimensional. For nontrivial dynamics, either $s_2$ or $s_3$ must be positive. 
Notice that although the underlying three- and four-wave interactions couple three and four photons, the dimension of the subspace $D=\min(s_2, s_3)+1$ can take any integer value. The smallest nontrivial problem is $D=2$, which requires a single qubit. 

The exact dynamics involves two fundamental frequency scales $g$ and $R$, from $H_T$ and $H_F$, respectively [Eqs.~\eqref{eq:HT} and~\eqref{eq:HF}]. When $g=0$, the dynamics is trivial because $H_F$ is diagonal: Under the influence of $H_F$ alone, the occupation of $|\phi_j\rangle$ remains unchanged, and the dynamics is a pure phase precession. To change occupation numbers, a nonzero $g$ is needed. Hence, for nontrivial dynamics, we can always normalize time by $\tau=|g|t$ and normalize the four-wave coupling by $\rho=R/|g|$. The Schr\"{o}dinger equation becomes $\textrm{i}\partial_\tau \mathbf{c} = \mathsfbi{H}\mathbf{c}$, where the vector $\mathbf{c}$ is the expansion coefficients such that $|\phi\rangle=\sum_{l=0}^{D-1} c_l|\phi_{k}\rangle$. The index $k=j_{\min}+l$, so $c_0$ is the probability amplitude that the seed wave is in the lowest occupied state, and $c_{D-1}$ is the probability amplitude that the seed wave is in the highest occupied state.
The normalized matrix elements $H_{k',k}=\frac{1}{|g|}\langle\phi_{k'}|H|\phi_k\rangle$ are nonzero only along the diagonal and first off-diagonal bands
\begin{subeqnarray}
    \label{eq:matrix}
    H_{k,k} &=& -\frac{1}{2}\rho k(k-1), \\
    H_{k-1,k} &=&e^{\textrm{i}\theta}\sqrt{k(s_2+1-k)(s_3-s_2+k)},     
\end{subeqnarray}
where $e^{\textrm{i}\theta} = \textrm{i} g/|g|$ and $H_{k,k-1}=H^*_{k-1,k}$.
Notice that for the amplitude $c_l$, the index $k=j_{\min}+l$ may be shifted. For example, when $s_2=3$ and $s_3=4$, we have $D=4$ and $j_{\min}=0$, so $k=l=0,1,2,3$. On the other hand, when $s_2=4$ and $s_3=3$, even though we still have $D=4$, now $j_{\min}=1$, so $k=l+1=1,2,3,4$.
In terms of $l$, we can rewrite $H_{k-1,k} =e^{\textrm{i}\theta}[l(D-l)(|s_3-s_2|+l)]^{1/2}$, which is symmetric under $s_2\leftrightarrow s_3$. 
The three-wave interaction is a phase-sensitive process because when $\varphi\ne\varphi'$, $|e^{\textrm{i}\varphi} + e^{\textrm{i}\varphi'}|<2$. Here, the phase $\varphi$ contains a contribution from $\theta$, as well as a dynamical phase, which accumulates in time due to the diagonal four-wave interaction. If the relative phase between two adjacent levels change, then the population transfer between them is reduced.

\begin{figure}
  \centering
  \includegraphics[width=0.75\textwidth]{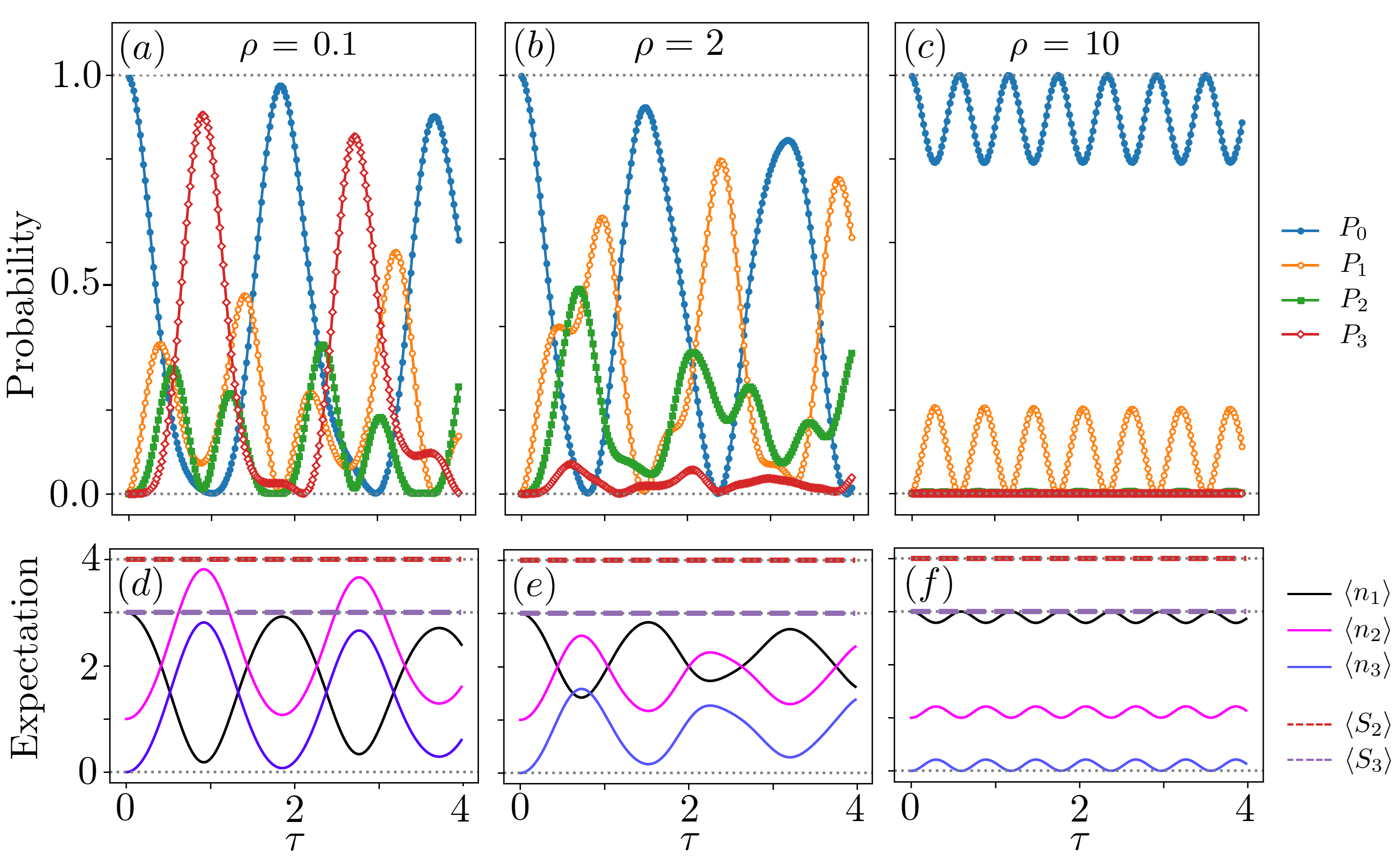}
  \caption{Exact dynamics of mixed three- and four-wave interaction problems in a $D=4$ dimensional Hilbert space with constants of motion $s_2=4$ and $s_3=3$. Starting from the ground state, the probability amplitudes $\mathbf{c}$ are evolved in time, and the occupation probabilities $P_l=|c_l|^2$ (\textit{a})-(\textit{c}), as well as the expected quanta in the three waves $\langle n_i\rangle$ (\textit{d})-(\textit{f}) are computed on a classical computer. When $\rho=R/|g|=0.1$, three-wave interaction dominates; when $\rho=2$, three- and four-wave interactions compete; when $\rho=10$, four-wave interaction dominates.}
\label{fig:exact}
\end{figure}

Since the Hamiltonian $\mathsfbi{H}$ is time independent, the exact dynamics is described by the unitary evolution operator $\mathsfbi{U}(\tau)=\exp(-\textrm{i}\mathsfbi{H}\tau)$, which can be obtained by direct diagonalization and exponentiation, at least for small problem sizes. For size $D=4$, the exact behaviors of three examples are shown in Fig.~\ref{fig:exact}. In all three examples, the coupling phase $\theta=0$ and the constants of motion are $s_2=4$ and $s_3=3$, which means $j_{\min}=1$ and $k=l+1$, so $|\phi_{k}\rangle=|3-l,1+l,l\rangle$. 
All examples start the time evolution from $c_0=1$ and $c_l=0$ for $l=1,2,3$ at $\tau=0$. This ground state of the computational basis is by no means the ground state of the dynamical system. In fact, at $\tau=0$, there are three photons in the pump wave $a_1$ and one photon in the seed wave $a_2$, while the plasma wave $a_3$ is initially unexcited.
At small $\tau$, the three-wave interaction consumes $a_1$ to produce $a_2$ and $a_3$, so the probabilities cascade from the ground state to higher states of the computational basis. At larger $\tau$, the pump is depleted, so the inverse process dominate, where $a_2$ and $a_3$ merge into $a_1$, and the probabilities cascade back to lower states. As can be seen from Fig.~\ref{fig:exact}, the upward and downward cascades of probabilities repeat, but the dynamics is not periodic.

The dynamics is controlled by the dimensionless parameter $\rho$. 
When $\rho\ll 1$, as shown in Fig.~\ref{fig:exact}(\textit{a}) and \ref{fig:exact}(\textit{d}), three-wave interaction dominates, which causes population transfer between the three waves. In this case, the much weaker four-wave interaction slowly accumulate phase mismatches that reduces the efficiency of population transfer, which is manifested by the decreasing oscillation amplitudes of $\langle n\rangle$ in Fig.~\ref{fig:exact}(\textit{d}). 
Here, the expected number of quanta in the three waves are calculated from the probability amplitudes $\mathbf{c}$ by
$\langle n_1 \rangle = \sum_l (s_2-j)|c_l|^2$, $\langle n_2 \rangle = \sum_l j|c_l|^2$, and $\langle n_3 \rangle = \sum_l (s_3-s_2+j)|c_l|^2$, 
where $j=j_{\text{min}}+l$, and the summation is over $l=0\dots,D-1$. From the above equations, it is clear that $\langle S_2\rangle=s_2$ and $\langle S_3\rangle=s_3$ are exact constants of motion, as marked by horizontal dashed lines in the lower panels of Fig.~\ref{fig:exact}. 
In the opposite limit $\rho\gg 1$, as shown in Fig.~\ref{fig:exact}(\textit{c}) and \ref{fig:exact}(\textit{f}), four-wave interaction dominates. In this case, the phases of different $|\phi_j\rangle$ precess at drastically different rates, which inhibits population transfer because three-wave interaction requires phase matching. In this example ($\rho=10$), the $0\leftrightarrow 1$ transfer is strongly suppressed, while the $1\leftrightarrow 2$ transfer, which accumulate phase mismatch at a greater rate, becomes nearly impossible.
Finally, in the intermediate case $\rho\sim 1$, as shown in Fig.~\ref{fig:exact}(\textit{b}) and \ref{fig:exact}(\textit{e}), three- and four-wave interactions compete, and the dynamics is more complicated. While four-wave interactions generate phase mismatches that suppress population transfer, three-wave interactions change the populations and affect how the phases are weighted. The intermediate cases are where simulations are most needed for predicting the behavior of the system.

The expectation value $\langle n (\tau) \rangle$ can be measured efficiently using $O(\log D)$ circuits. Due to action conservation, the three expectation values $\langle n_1 \rangle = s_2 - \langle n_2 \rangle$, $\langle n_2 \rangle = j_{\min} + \langle \mathsfbi{O} \rangle$, and $\langle n_3 \rangle = s_3 - s_2 + \langle n_2 \rangle$ are derivable from
\begin{equation}
    \label{eq:expectation}
    \langle \mathsfbi{O} \rangle = \sum_{l=0}^{2^n-1} l |c_l|^2 = \frac{2^n-1}{2} - \sum_{j=1}^n 2^{n-1-j} \langle \phi|\mathsfbi{Z}_j|\phi\rangle,
\end{equation}
where we embed the $D$ level system into the Hilbert space of $n=\lceil \log_2(D) \rceil$ qubits. We identify $|l\rangle = |\phi_{j_{\min}+l}\rangle$ and map this state to $|q_1\dots q_n\rangle$ when $l=(q_1\dots q_n)_2$, where $q=0$ or $1$ and $(q_1\dots q_n)_2 = 2^{n-1}q_1  + \dots 2^0 q_n$ is the binary representation of $l$. When $N=2^n>D$, we pad the state vector with zeros, namely, we set $c_l=0$ for $l\ge D$. On the right-hand side of Eq.~(\ref{eq:expectation}), $|\phi\rangle = \sum_l c_l |l\rangle$ is the quantum state, and $\mathsfbi{Z}_j$ is the Pauli $\sigma_z$ gate acting on the $j$-th qubit. By measuring $n$ single-qubit Pauli strings $\langle \mathsfbi{Z}_j\rangle$ for $j=1,\dots, n$, Eq.~(\ref{eq:expectation}) allows us to obtain $\langle \mathsfbi{O} \rangle$ using $n=O(\log D)$ measurement circuits with $n$ steps of post processing. 
To see why Eq.~(\ref{eq:expectation}) holds, we start from $\mathsfbi{O}=\sum_l l |l\rangle\langle l |$. In the computational basis, $\mathsfbi{O}=\textrm{diag}(0,1,\dots, N-1)$ is a diagonal matrix, which can be expanded in Pauli basis as $\mathsfbi{O}=\sum_l r_l \sigma^n_l$ where $\sigma^n_l$ is a tensor product of $n$ Pauli matrices. Explicitly, $\sigma^n_{(q_1\dots q_n)_2}:=\bigotimes_{i=1}^n \sigma_{3q_i}$, where $\sigma_{3q}=\mathsfbi{I}$ when $q=0$ and $\sigma_{3q}=\mathsfbi{Z}$ when $q=1$. 
Using the fact that $\text{tr}(\sigma^n_i\sigma^n_j)=2^n\delta_{ij}$, where $\delta_{ij}$ is the Kronecker delta, the expansion coefficient equals $r_l=2^{-n}\text{tr}(\sigma^n_l \mathsfbi{O})$. 
To calculate the trace, we need diagonal elements of $\sigma^n_l$, and we denote its $k$-th diagonal element by $[\sigma^n_l]_k$. By induction, one can show that $[\sigma^n_{(q_1\dots q_n)_2}]_{(b_1\dots b_n)_2} = \prod_{i=1}^n (-1)^{q_ib_i}$. Then, $\text{tr}(\sigma^n_{(q_1\dots q_n)_2} \mathsfbi{O}) = \sum_k [\sigma^n_{(q_1\dots q_n)_2}]_k [\mathsfbi{O}]_k = \sum_{b_1,\dots, b_n=0,1} [\sigma^n_{(q_1\dots q_n)_2}]_{(b_1\dots b_n)_2} (b_1\dots b_n)_2 = \sum_{b_1,\dots, b_n=0,1} \prod_{i=1}^n (-1)^{q_ib_i} \sum_{j=1}^n 2^{n-j} b_j = \sum_{j=1}^n 2^{n-j} \xi_j$.
To calculate the sum $\xi_j=\sum_{b_1,\dots, b_n=0,1} \prod_{i=1}^n (-1)^{q_ib_i} b_j$, one can show by induction that if $q_i=0\; \forall i\ne j$, $\xi_j= 2^{n-1}(-1)^{q_j}$, otherwise $\xi_j=0$.  
Consequently, only $n+1$ traces are nonzero. 
The first nonzero trace is when $q_i=0$ for all $i$, in which case $\sigma^n_{(0\dots 0)_2} = \mathsfbi{I}$ and $\text{tr}(\mathsfbi{O})=2^{n-1}(2^n-1)$. Second, when $q_i=\delta_{ij}$ for a given $j$, $\sigma^n_{(0\dots 1_j \dots 0)_2} = \mathsfbi{Z}_j$ is a single-qubit gate and $\text{tr}(\mathsfbi{Z}_j\mathsfbi{O})=-2^{n-1} 2^{n-j}$. When more than one $q$ is nonzero, the trace is zero. 
Therefore, we obtain $\mathsfbi{O}=2^{-1}(2^n-1)\mathsfbi{I}-\sum_{j=1}^n 2^{n-1-j}\mathsfbi{Z}_j$, from which Eq.~(\ref{eq:expectation}) follows.

The exact quantum dynamics can be simulated efficiently using quantum Hamiltonian simulations \citep{berry2007efficient,berry2014exponential,low2017optimal}, exploiting the fact that our Hamiltonian matrix is 3-sparse.
For example, the qubitization algorithm of \citet{low2017optimal}, requires $O(1)$ ancilla qubits in addition to $O(\log D)$ qubits that encode the action basis.
The query complexity of the qubitization algorithm, namely, the number of terms one need to keep in the Jacobi-Anger expansion, is $O(\tau\|H\|_{\max} + \log(1/\epsilon)/\log\log(1/\epsilon))$, where $\epsilon$ is the desired precision and $\|H\|_{\max}$ is the matrix element with the largest absolute value. 
For plasma pulse compression, one is typically interested in attaining the maximum seed wave intensity. As shown in Fig.~\ref{fig:exact}, the first peak of $\langle n_2\rangle$ tends to be the highest. So, it is usually sufficient to simulate the dynamics for the first few oscillations. In other words, the maximum simulation time of interest is $\tau=O(1/|\lambda|_{\max})$, where $|\lambda|_{\max}$ is the eigenvalue of $H$ that has the largest absolute value. 
Therefore, the query complexity for the pulse compression problem is $O(\|H\|_{\max}/|\lambda|_{\max})$. 
In other problems, such as simulating the chaotic regime of wave-wave interactions, one might be interested in long-time dynamics, in which case the complexity may be higher.
To estimate $\|H\|_{\max}$, using Eq.~(\ref{eq:matrix}), the diagonal elements attain the maximum at the largest $k$ value, which is $k=s_2$. When $s_2\rightarrow\infty$, $\|H_{k,k}\|_{\max}\simeq\frac{1}{2}\rho s_2^2$. 
For off-diagonal elements, we express $H_{k,k-1}$ in terms of $l$.
When $s_3\rightarrow\infty$ at fixed $s_2$, or $s_2\rightarrow\infty$ at fixed $s_3$, the maximum is attained at $l\simeq D/2$, where 
$\|H_{k,k-1}\|_{\max}\simeq (D/2)\sqrt{|s_2-s_3|+D/2}$.
When $s_2=s_3\rightarrow\infty$, the maximum is attained at $l\simeq 2D/3$ where $\|H_{k,k-1}\|_{\max}\simeq2(D/3)^{3/2}$. 
Although $\|H\|_{\max}$ grows with $D$, the complexity remains $O(1)$ because $|\lambda|_{\max}$ also grows with $D$ at a similar rate. 
To estimate the eigenvalues, consider two limits. 
(i) Consider the limit $\rho\rightarrow\infty$. 
When $s_2> s_3$, because $j_{\min}>0$, diagonal matrix elements dominate, and the eigenvalues are $\lambda_l\simeq H_{k,k}\rightarrow-\infty$, for $l=0,\dots,D-1$. 
When $s_2\le s_3$, because $j_{\min}=0$, the first two diagonal elements are zero. The eigenvalues are $\lambda_0\simeq|H_{0,1}|, \lambda_1\simeq-|H_{0,1}|$ and $\lambda_l\simeq H_{k,k}$ for $l=2,\dots,D-1$, where $|H_{0,1}|^2=(D-1)(|s_2-s_3|+1)\ll \rho$. 
In both cases, $|\lambda|_{\max} \simeq \|H\|_{\max}$.
(ii) Consider the limit $\rho\rightarrow 0$. Due to the absence of diagonal elements, the eigenvalues are either $0$ or appear in $\pm\lambda_i$ pairs, where $i=1,\dots, \lfloor D/2\rfloor$. From the characteristic polynomial, $\sum_{i=1}^{\lfloor D/2\rfloor} \lambda_i^2 = \sum_{l=1}^{D-1} |H_{k-1,k}|^2=\frac{1}{12}D(D^2-1)(D + 2|s_2-s_3|)$. 
Because $\sum_{i=1}^{\lfloor D/2\rfloor} \lambda_i^2\le \lfloor D/2\rfloor |\lambda|^2_{\max}$, we obtain a lower bound $|\lambda|_{\max}\gtrsim [\frac{1}{6}(D^2-1)(D+2|s_2-s_3|)]^{1/2}$. Compare with expressions for $\|H_{k,k-1}\|_{\max}$, we see $\|H\|_{\max} = O(|\lambda|_{\max})$ when $D\rightarrow\infty$.
For both limits, as well as for intermediate $\rho$, the complexity of quantum Hamiltonian simulation is $O(\|H\|_{\max}/|\lambda|_{\max})=O(1)$, so the pulse compression problem can be simulated efficiently.    

\section{Implementing exact dynamics with error mitigation\label{sec:exact}}
The classical three- and four-wave interaction problems, when restricted to the temporal case with no spatial nonuniformity, are in fact not difficult to solve on classical computers.
However, the quantum problem, which becomes important, for example, when the pump is in a squeezed state, become challenging for classical computers if the number of photons is large. The exact diagonalization of the 3-sparse $D\times D$ Hamiltonian takes $O(D)$ steps and $O(D^2)$ memory. 
For joule-class lasers typically used in plasma physics, the number of photons is on the order of $10^{19}$, so exactly solving the quantum problem on classical computers is likely challenging. 
On the other hand, solving the quantum problem on future quantum computers will be efficient, which requires $O(\log D)$ qubits and $O(1)$ complexity. 
However, fault-tolerant quantum computers are not yet available. To push the limit of current devices and identify avenues for near-future improvements, we perform experiments on a superconducting device, using product-formula algorithms to approximate the exact dynamics.

We perform two-qubit experiments on Rigetti's Aspen-M-3 processor \citep{caldwell2018parametrically}, which is a superconducting device with multiple transmon qubits at a fixed topology with hardwired qubit-qubit couplings. The device is routinely calibrated to support single-qubit gates, as well as two-qubit gates like CZ and parametric XY($\theta$) gates. 
%
%
Each experiment is specified as a sequence of unitary operations, and each 4-by-4 unitary matrix is decomposed using Cartan decomposition into at most three two-qubit SQISW gates, sandwiched between single-qubit gates \citep{huang_quantum_2023}. The total gate sequence is executed on the hardware with the device initialized in the ground state. At the end of the gate sequence, the states of the two qubits are measured. The whole process of an experiment takes a few microseconds to run on hardware, with the overall time being dominated by a passive reset delay. We repeat each experiment for $M=50,000$ times to accumulate statistics for the final states, so that the shot noise, which scales as $O(1/\sqrt{M})\sim 0.4\%$, is small compared to other sources of errors. At the end of $M$ repeated experiments, we obtain a single data point along the time history of the evolution. 
Because projective measurement destroys quantum states, to obtain the next point along the time history, the simulation has to restart from the beginning in the form of a different experiment, which has its own sequence of unitary operations and is repeated another $M$ times.

\subsection{Implementation on a superconducting device}
\begin{figure}
  \centering
  \includegraphics[width=1.0\textwidth]{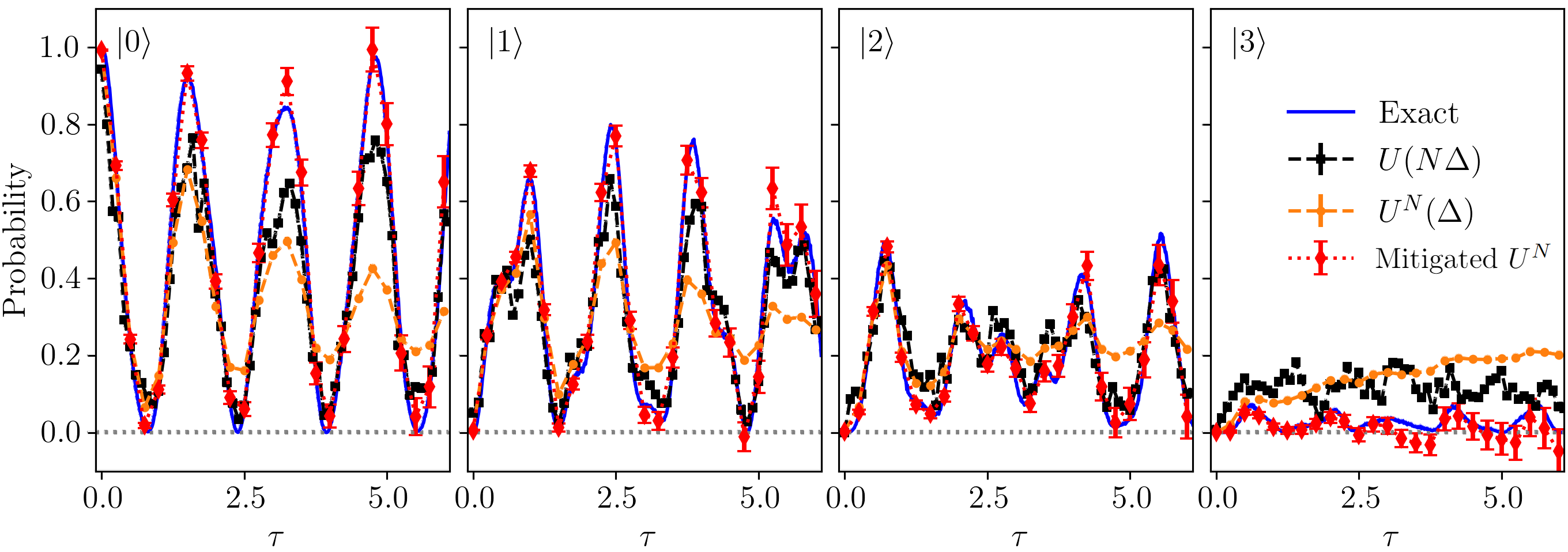}
  \caption{Occupation probabilities in a test problem with parameters $\rho=2$, $\theta=0$, $s_2=4$, and $s_3=3$. The blue lines are exact solutions from a classical computer, and the colored symbols with error bars are results from the quantum device. When asked to enact the final unitary (black), the device performance is acceptable but not ideal. However, when asked to perform time evolution (orange), results on the device degrade to noise level after a few oscillations. The results are significantly improved using error mitigation techniques (red), but the error bars grow exponentially.}
\label{fig:hardware}
\end{figure}

As the first test of the quantum device, we use it to enact the exact unitary operator. In Fig.~\ref{fig:hardware}, the solid blue lines are the exact occupation probabilities of the four states in our computational basis, which are computed using exact exponentiation on classical computers. The test problem uses parameters $\rho=2$, $\theta=0$, $s_2=4$, and $s_3=3$, which are identical to the middle panel of Fig.~\ref{fig:exact}. The exact solutions serve as references for results on the quantum device.
This first test is the simplest task that a quantum hardware can perform: For each time $\tau=N\Delta$, we compute the unitary exactly on a classical computer. The sequence of unitary operations for this experiment is thus constituted of just a single unitary, $\mathsfbi{U}(N\Delta)$, and the results are shown in Fig.~\ref{fig:hardware} as the black dashed lines. As can be seen from the figure, even when enacting a single dense unitary on the device, the fidelity is far from perfect. The important point is that because each time step uses the same gate depth, the performance does not degrade with $\tau$, except perhaps at the very beginning where most population remains in the ground state and the unitary is near identity. In this test, because the gate sequence is so short, decoherence is not a leading cause of infidelity. Instead, most infidelity comes from coherent gate errors, in the sense that each gate realizes a slightly different unitary than what is intended.

In this simplest test, another source of error is readout, for which we have already corrected using an iterative Bayesian unfolding technique \citep{nachman2020unfolding}. On the device level, the dispersive readout process is a scattering experiment, where a microwave pulse, whose frequency is off-resonant from the qubit transition frequency, is injected to interact with the qubit. The amplitude and phase of the returned microwave pulse depends on the state of the qubit, and therefore enables a measurement of the qubit states. During the readout process, the state of the qubit may change, for example, due to intrinsic decoherence or induced interactions from the readout pulse. As a consequence, the true probability $p_i$ that qubits are in state $i$ may differ from the measured probability $m_i$. The vectors $\mathbf{m}$ and $\mathbf{p}$ are related by a response matrix $\mathsfbi{R}$ by $\mathbf{m}=\mathsfbi{R}\mathbf{p}$, where $\mathsfbi{R}$ is also called the confusion matrix, which is ideally close to the identity matrix. 
In practice, the matrix $\mathsfbi{R}$ is constructed during calibrations, which prepare known quantum states using single-qubit gates, assuming they are ideal, and then immediately measure the qubit states. Due to statistical noise, directly inverting the matrix to find $\mathbf{p}=\mathsfbi{R}^{-1}\mathbf{m}$ often leads to artifacts, including $p\notin[0,1]$ and $\sum_i p_i\ne 1$. These artifacts can be removed using iterative Bayesian unfolding $p^{n+1}_i=\sum_j (m_jR_{ji}p_i^n/\sum_kR_{jk}p_k^n)$, where $n$ is the iteration number. When the iteration converges, the denominator $\sum_kR_{jk}p_k=m_j$ cancels with the numerator. Because the marginalized probability $\sum_j R_{ji}=1$ for all $i$, $p^{n+1}_i=p^n_i$ converges to a stationary true value. In practice, starting from the initial guess $\mathbf{p}^0=\mathbf{m}$, a few iterations are usually sufficient. 
Without readout error mitigation, results differ only slightly from the black dashed lines in Fig.~\ref{fig:hardware}, which suggests that the leading error is not readout error but rather coherent gate errors.

As the second test, we perform time evolution using the exact unitary $\mathsfbi{U}(\Delta)$, and the results are shown by the orange dashed lines in Fig.~\ref{fig:hardware}. In this set of experiments, $\mathsfbi{U}(\Delta)$ is compiled to native gates, and the gate sequence is repeated $N$ times to enact $\mathsfbi{U}^N(\Delta)$. Because of the repetition, as $\tau=N\Delta$ increases, the gate depth increases linearly. The accumulation of errors leads to a degradation of fidelity, as can be seen from Fig.~\ref{fig:hardware}. The oscillation amplitudes decrease and $p(\tau)$ deviates further from the true solution as $\tau$ increases. At even larger $\tau$ values, the quantum states become fully scrambled, so $p\rightarrow 1/4$ approaches the fully mixed value for the four quantum states. Because $\mathsfbi{U}^N(\Delta)$ has a larger depth than $\mathsfbi{U}(N\Delta)$, the device performs worse in this test (orange lines) than in the previous test (black lines) as expected.
The $\mathsfbi{U}^N(\Delta)$ results improve noticeably from \citet{shi2021simulating} primarily because of the SQISW gate, which has significantly shorter duration and higher fidelity than two-qubit gate used in our previous work.

\subsection{Improving results with error suppression and mitigation\label{sec:mitigation}}
Because product formulas require even larger gate depth, we need to improve the results for the exact unitary before moving on to the next test. 
The dominant source of error on the quantum processor are two-qubit (2Q) gate errors, which are typically an order of magnitude larger than single-qubit (1Q) gate errors. On the Aspen architecture, this is not only due to a significantly longer 2Q gate time, but also because activating the 2Q gate requires tuning one of the qubits away from its optimal operating point so the qubit becomes more sensitive to flux noise ~\citep{didier_flux_2019}, which leads to a higher dephasing rate for the qubit.

We take a multi-pronged approach to minimizing 2Q gate errors. First, the Aspen-M-3 chip used in our experiment has $\sim$200 calibrated 2Q gates available. We require only one of these for this circuit and are thus able to select high-performing candidates based on the reported fidelities. In our experiment, we perform this selection manually, but this optimization can be performed by compilers and is often referred to as \textit{addressing}.
Secondly, the Aspen chip offers both native CZ and XY($\theta$) gates, thus providing a choice of how to express our problem unitary. We observe that the XY family is particularly expressive ~\citep{peterson_fixed-depth_2020}, allowing the expression of our target unitary using two XY($\pi /2)$ gates and single-qubit gates ~\citep{huang_quantum_2023}. The parametric XY($\theta$) gate is composed of two pulses with a swap angle of $\pi/2$ while the variable angle is achieved by a phase-shift between the pulses ~\citep{abrams_implementation_2020}. However, we find that by using SQISW as our native gate, we can discard the second pulse from the XY gate, cutting the duration in half and reducing the 2Q gate error by around 40 percent. Our native 2Q gate is thus a 64-ns SQISW which is combined with single-qubit rotations to produce highly expressive native cycles. Our target unitary is expressible in either two or three of these clock cycles depending on parameters of the problem. Our combined approach to native gate selection, which takes into account both the fidelity and expressiveness of the native 2Q gates, allows us to achieve sufficiently low error rates so that it is possible to reach relatively large depths.

While our calibration and gate selection techniques suppress errors, setting the fundamental performance ceiling of the circuit, a family of error mitigation techniques allows us to tailor the noise, and recover unbiased estimates of observables.
As mentioned earlier,  a major source of error is coherent gate error. Coherent errors may be the result of either control errors, which should have ideally been calibrated away, or crosstalk, which are coherent errors caused by the state of nearby idling spectator qubits. Such errors can be suppressed using dynamical decoupling \citep{tripathi2021suppression,Zeyuan:22, evert2024syncopated}. We do not attempt this here because our circuit does not have significant idle periods. Finally, we may suffer from unintended control errors from neighbouring control signals. All three classes of errors are major contributors to infidelity on current devices. 
Moreover, while long-term changes, such as temperature drifts that affect the control electronics, can be removed by routine calibrations, current devices also suffer from short-term and unpredictable changes. 
For example, chemical residues from fabrication process or material defects can interact with the qubits, which changes qubit frequencies and coherence times \citep{PhysRevLett.121.090502,cho2023direct,muller2019towards}. As another example, when the device is struck by cosmic rays, whose energy is often higher than the superconducting gap, Cooper pairs are broken, creating quaisparticles, and thus, changing the qubit frequency \citep{vepsalainen2020impact}. 
Due to these changes, a control pulse that is calibrated to enact a specific unitary for the original set of qubit parameters will now deviate from the intended unitary operation.
The resultant coherent gate errors are particularly damaging because their behaviors are difficult to predict. Rather than adding up systematic errors in a simple way, quantum interference may cause the errors to transiently disappear, only to reemerge at a later time. 
In comparison, errors due to decoherence, or more specifically due to depolarizing noise, behave in a much simpler and more predictable way: the depolarizing errors simply lead to an exponential decay of the Bloch vector, which can be easily modeled once the decay exponent is measured.

To mitigate coherent errors, we convert them into stochastic Pauli errors using a random compilation technique \citep{Wallman16,hashim2020randomized}. The technique exploits the fact that the decomposition of a target unitary into elementary gates is not unique. 
Suppose one has found a particular decomposition $\mathsfbi{U}=(\prod_{k=1}^K \mathsfbi{C}_{k}\mathsfbi{G}_k)\mathsfbi{C}_{0}$, where $\mathsfbi{C}$ are 1Q gates and $\mathsfbi{G}$ are 2Q gates. Then, a different but equivalent decomposition can be constructed in two steps. 
First, the 1Q gates are replaced by $\tilde{\mathsfbi{C}}_{k}=\mathsfbi{T}_{k+1}\mathsfbi{C}_{k} \mathsfbi{T}^c_{k}$ for $k=0,\dots,K$, where $\mathsfbi{T}_{k+1}$ is chosen at random from a subgroup of $\mathsfbi{C}$ called the twirling group. 
An important property of the twirling group is that $\mathsfbi{T}\mathsfbi{G}=\mathsfbi{G}\mathsfbi{T}'$. In other words, a 1Q gate in $\mathsfbi{T}$ can be commuted across 2Q gates to become another 1Q gate. 
Due to this property, once $\mathsfbi{T}_k$ is chosen, if one takes $\mathsfbi{T}^c_k=\mathsfbi{G}_k\mathsfbi{T}^\dagger_k\mathsfbi{G}^\dagger_k$ for $k=1,\dots,K$, then $\mathsfbi{T}^c_k\mathsfbi{G}_k\mathsfbi{T}_k=\mathsfbi{G}_k$. 
Taking $\mathsfbi{T}^c_0=\mathsfbi{I}$ to be the identity, one thus finds another decomposition $\mathsfbi{U}=\mathsfbi{T}^\dagger_{K+1}(\prod_{k=1}^K \tilde{\mathsfbi{C}}_{k}\mathsfbi{G}_k)\mathsfbi{C}_{0}$.
Second, to reduce unnecessary 1Q gate depth, the new 1Q operations $\tilde{\mathsfbi{C}}_{k}$, for $k=1,\dots,K-1$, and $\mathsfbi{T}^\dagger_{K+1}\tilde{\mathsfbi{C}}_{K}$ are compressed and simplified to elementary 1Q gates. 
In this paper, the 2Q gate we use is the SQISW gate $\mathsfbi{G}=\exp[-i\frac{\pi}{8}(\mathsfbi{X}\otimes\mathsfbi{X} +\mathsfbi{Y}\otimes\mathsfbi{Y})]$, where $\mathsfbi{X}$ and $\mathsfbi{Y}$ are Pauli matrices.
Notice that the twirling does not directly touch the 2Q gates, which are hard to implement on hardware. The randomness of the compilation is introduced solely from the intermediate 1Q gates, which are easy to adjust. 
\citet{Wallman16} shows that if the Pauli twirling gates are chosen independently and if hardware Pauli errors are also independent, then random compilation transforms any gate errors into stochastic Pauli errors.
In other words, as shown in Appendix~\ref{appA}, suppose the errors of a quantum channel, when represented by the Pauli-transfer matrix, have off-diagonal components before twirling. Then, after twirling, the errors become purely diagonal, which means coherent interference of errors is removed. Because our native two-qubit gate is a non-Clifford gate, we cannot apply full Pauli twirling. Rather, we use a pseudo-twirling technique which tailors a smaller subset of coherent errors using the group of single-qubit rotations which can be successfully inverted by 1Q gates. This twirling group is less powerful than Pauli twirling, but still tailors the noise effectively in most situations. The twirling is performed using the TrueQ software library \citep{trueq}.
In our experiments, we construct the logical circuit and compute 50 random compilations. Each compilation has an identical pulse schedule, and thus an identical noise model. The randomization of 1Q gates is performed by updating angles of our virtual $\mathsfbi{Z}$ gates. Such updates can be made with high efficiency, allowing a large number of randomizations of the circuit to be executed in quick succession.

The final step of error mitigation is to compensate for the suppression of observables using a rescaling technique \citep{Ville22}. After twirling of a unitary operation $\mathsfbi{U}$, the noise channel becomes approximately $\mathcal{E}(\rho)=\sum\lambda_P \mathsfbi{P}\rho \mathsfbi{P}^\dagger$, where the summation is over all tensor products of 1Q Pauli operators $\mathsfbi{P}$. The coefficient $\lambda_P$, called the Pauli decay constant, is specific to the Pauli operator $\mathsfbi{P}$ but is independent of the unitary $\mathsfbi{U}$ that is being performed. Because the $\lambda_P$'s are bounded by their mean $\bar{\lambda}$ as $2\bar{\lambda}-1\le\lambda_P\le 1$, \citet{Ville22} propose to use a single $\bar{\lambda}$ value to correct for all Pauli errors. This approximation becomes exact when the errors are fully depolarizing, which means that all Pauli channels decay in the same way. In this case, the measured expectation value $\tilde{E}$ for any $E$ of interest is given by a simple rescaling $\tilde{E}=\bar{\lambda}E$, because the length of the Bloch vector, which measures the purity of the state, shrinks by $\bar{\lambda}$. 
For example, in our two-qubit problem, the expected occupation of a state beyond its fully mixed value is $(\tilde{p}-\frac{1}{4})=\bar{\lambda}(p-\frac{1}{4})$ , where $p$ is the occupation probability for a pure state after one unitary operation. Then, after $N$ unitary operations, we can purify the probability by a rescaling $p=\frac{1}{4}+\bar{\lambda}^{-N}(\tilde{p}-\frac{1}{4})$.
In other words, after measuring the probability of a state, we subtract the noise $\frac{1}{4}$ and amplify the remaining signal exponentially by a rescaling factor $(1/\bar{\lambda})^{N}$. We provide justifications for using the simple rescaling technique in Appendix~\ref{appA}. Notice that while amplifying the signal, this purification procedure also amplifies statistical error bars exponentially.

In practice, we estimate the mean value of Pauli decay constants $\bar{\lambda}$ using cycle error reconstruction \citep{erhard2019characterizing, carignan-dugas_error_2023}. Cycle error reconstruction is a technique for measuring the Pauli infidelities of a dressed cycle with multiplicative precision. A Pauli-twirled cycle results in a noise channel which is diagonal in the Pauli basis. After the qubits are prepared in an initial state, the cycle of interest is repeated $M$ times and inverted to create an identity. By repeating the procedure at different values of $M$, we can we extract a state-preparation and measurement (SPAM) robust error rate for the initial state. The ensemble of error rates allow us to reconstruct the Pauli error rates of the cycle and their orbital averages, as shown in Appendix~\ref{appA}. Together, this set of measurements provides not only an overall measurement of the fidelity, but structured information about the error channel which can be used in more advanced error mitigation techniques. The technique we use is scalable with the number of qubits, provided that the noise is sparse, which is a reasonable assumption in superconducting qubit architectures. Empirically, the result converges with the square root of the sampling size. 
In our case, the gate of interest is the SQISW gate. For a system of two qubits, it is affordable to perform cycle error reconstruction over all possible Pauli channels. The SQISW is a non-Clifford gate, meaning that Pauli twirls cannot be inverted by a following pair of 1Q rotations. However, for the purposes of characterization, this is not a problem. The inversions are propagated to the end of the benchmarking circuit, and then inverted with a final SQISW gate.  This procedure of measuring the purification constant $\bar{\lambda}$ 
is performed each time we run a batch of experiments on the hardware. 

After performing the above error mitigation steps, the hardware results for our test problem are shown in Fig.~\ref{fig:hardware} by the dotted red lines. The mitigated results of the second test now closely tracks the exact solutions, and performs even better than the first test (black lines), which does not use any mitigation.
While the mitigation significantly improves the signals, without noticeable increasing the hardware overhead, the price we pay is exponentially growing error bars. At even larger simulation depth, the error bars will become comparable to the signals, beyond which the simulations need to stop. 


\section{Testing product formulas with limited quantum resources\label{sec:trotter}}
With sufficient simulation depth, we can now test the next level of quantum simulations, without assuming that the exact unitary is known. Predicting the exact dynamics on classical computers requires exact exponentiation of $\mathsfbi{H}$, which becomes expensive for large problem sizes. As a prototypical step of quantum simulation, we split $\mathsfbi{H}=\mathsfbi{H}_T+\rho\mathsfbi{H}_F$ into two non-commuting terms, where the normalized four-wave coupling $\rho$ is pulled out from the definition of $\mathsfbi{H}_F$. 
On future quantum computers, if the total Hamiltonian cannot be simulated efficiently, but its components can, then product formulas may be employed to approximate the exact dynamics.
In our case, $\mathsfbi{H}_F$ is diagonal, so it is trivial to exponentiate on classical computers; $\mathsfbi{H}_T$ is tridiagonal with zero diagonal elements, so its spectrum appears in $\pm\lambda$ pairs. Otherwise, $\mathsfbi{H}$ is not intrinsically more difficult to exponentiate than $\mathsfbi{H}_T$ on classical computers. We do not expect a future quantum computer to benefit from separating $\mathsfbi{H}$ into $\mathsfbi{H}_T$ and $\mathsfbi{H}_F$ when performing quantum Hamiltonian simulations. Nevertheless, we perform the separation in order to test the performance of product formulas on current quantum devices. 

\subsection{Product formulas}
For given problem parameters, we compile unitary matrices $\mathsfbi{U}_T(\tau)=\exp(-\textrm{i}\mathsfbi{H}_T\tau)$ and $\mathsfbi{U}_F(\tau)=\exp(-\textrm{i}\mathsfbi{H}_F\tau)$ to native gates using a Cartan decomposition \citep{smith2020open}. Then, we use $\mathsfbi{U}_T$ and $\mathsfbi{U}_F$ to approximate the exact $\mathsfbi{U}$. To first order, we approximate a single step with time step size $\tau=\Delta$ using the formula
\begin{equation}
    \label{eq:U1}
    \mathsfbi{U}_1(\Delta) = \mathsfbi{U}_T(\Delta) \mathsfbi{U}_F(\rho\Delta),
\end{equation}
where the normalized four-wave coupling $\rho$ appears as a scaling of time, which is convenient on quantum computing platforms that support parametric compiling.
The error per step of the first order formula is $O(\Delta^2)$, whose prefactor is proportional to the norm of the commutator $[\mathsfbi{H}_T, \mathsfbi{H}_F]$.
To second order, we use a symmetric formula
\begin{equation}
    \label{eq:U2}
    \mathsfbi{U}_2(\Delta) = \mathsfbi{U}_T\Big(\frac{\Delta}{2}\Big) \mathsfbi{U}_F(\rho\Delta) \mathsfbi{U}_T\Big(\frac{\Delta}{2}\Big),
\end{equation}
whose error per step is $O(\Delta^3)$. One could instead place $\mathsfbi{U}_F$ on the outside. With either placement, when repeating $\mathsfbi{U}_2$, the adjacent $\mathsfbi{U}$ of the same kind can be merged to reduce the required number of operations. 
For example, using Eq.~(\ref{eq:U2}), we can simplify
$\mathsfbi{U}^2_2(\Delta)=\mathsfbi{U}_T(\Delta/2)\mathsfbi{U}_F(\rho\Delta)\mathsfbi{U}_T(\Delta)\mathsfbi{U}_F(\rho\Delta)\mathsfbi{U}_T(\Delta/2)$. Implementing this unitary sequence thus requires compiling three different unitary matrices $\mathsfbi{U}_T(\Delta/2)$, $\mathsfbi{U}_T(\Delta)$, and $\mathsfbi{U}_F(\rho\Delta)$. In a similar spirit, higher order product formulas require more types of unitary operations. For example, to third order, we use the product formula
\begin{equation}
    \label{eq:U3}
    \mathsfbi{U}_3(\Delta) = 
    \mathsfbi{U}_T\Big(\frac{7\Delta}{24}\Big) 
    \mathsfbi{U}_F\Big(\frac{2\rho\Delta}{3}\Big)
    \mathsfbi{U}_T\Big(\frac{3\Delta}{4}\Big) 
    \mathsfbi{U}_F\Big(-\frac{2\rho\Delta}{3}\Big)
    \mathsfbi{U}_T\Big(-\frac{\Delta}{24}\Big)     
    \mathsfbi{U}_F(\rho\Delta),
\end{equation}
whose error per step is $O(\Delta^4)$. Notice that two of the six unitary operations above are evolving backward in time. This feature is common for high order product formulas: In order to cancel higher order errors, more operations are needed and the time step sizes become increasingly constrained such that negative values become necessary. 
To achieve even higher order approximations, we use Suzuki's symmetric recurrence formula \citep{suzuki1990fractal}, which allows construction of $\mathsfbi{U}_{k+2}$ from $\mathsfbi{U}_k$ by
\begin{equation}
    \label{eq:Suzuki}
    \mathsfbi{U}_{k+2}(\Delta) = 
    \mathsfbi{U}^2_k(p\Delta) 
    \mathsfbi{U}_k[(1-4p)\Delta]
    \mathsfbi{U}^2_k(p\Delta), 
\end{equation}
where $p=1/(4-4^{1/(k+1)})>1/4$, so the middle step is an evolution backward in time. Suzuki's formula has a self-similar structure once it is fully expanded in terms of elementary $\mathsfbi{U}_T$ and $\mathsfbi{U}_F$ operations. As the order $k$ increases, $\mathsfbi{U}_k$ becomes increasingly accurate with $O(\Delta^{k+1})$ errors per step, but the required number of elementary operations increases exponentially as $O(5^{k/2-1})$, which incurs a significant computational cost. Thus, in practice, high order Suzuki formulas are rarely used and low-order formulas offer the best compromise of accuracy versus speed. 

\begin{figure}
  \centering
  \includegraphics[width=1.0\textwidth]{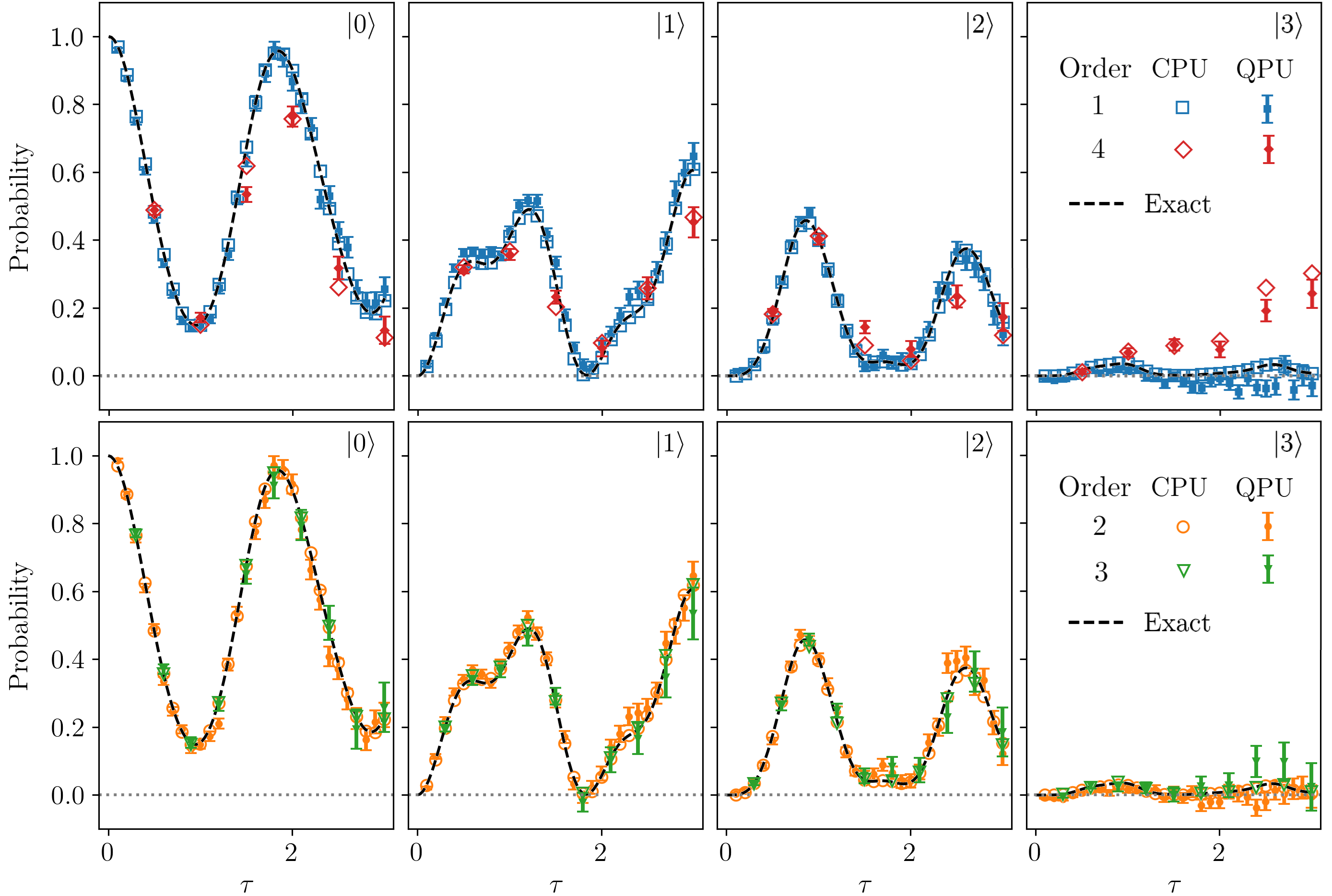}
  \caption{At a fixed gate depth, different product formulas are used for a test problem with parameters $\rho=4$, $\theta=0$, $s_2=3$, and $s_3=3$. For clarity, results from four cases are split in two separate panels. The exact results are obtained by exponentiating the total Hamiltonian on a classical computer. Results of product formulas on a classical computer are shown by open symbols, and results obtained on quantum devices are shown by solid symbols with error bars.
  Because higher order algorithms require more gates per step, for a fixed gate depth, they must use larger time step sizes to reach the same targeted final time $\tau_f=3$, leading to larger discretization errors. 
  }
\label{fig:trotter}
\end{figure}

With error mitigation techniques, we are able to run experiments on the quantum device for up to about two hundred 2Q gates. The gate depth is deep enough that we can begin to compare results of different product formula algorithms, which are shown in Fig.~\ref{fig:trotter}. 
In this set of experiments, different product formulas are employed to reach the same final simulation time $\tau_f=3$ at a fixed gate depth. We choose not to exhaust the maximum gate depth such that error bars at the final time remain small.

With a fixed gate budget, because lower order algorithms [Eqs.~(\ref{eq:U1}) and (\ref{eq:U2})] require fewer unitary operations per step, they can afford to use smaller time step sizes. In Fig.~\ref{fig:trotter}, both first-order (blue) and second-order (orange) algorithms yield results that match closely with the exact solutions (dashed lines), which are obtained on a classical computer by exact diagonalization and exponentiation of the total Hamiltonian matrix. Moreover, results on the quantum device (solid symbols) closely match results when the same product formula is used on a classical computer (open symbols).

In contrast, higher order algorithms require significantly more unitary operations per step, and thus can only afford to use a much larger $\Delta$ for a fixed total gate depth. 
At third order [Eq.~(\ref{eq:U3})], the time resolution is still sufficient to capture the oscillatory dynamics, and results on hardware are close to expected results (Fig.~\ref{fig:trotter}, green symbols). 
However, at fourth order [Eq.~(\ref{eq:Suzuki})], the coarse time step leads to a large discretization error. Although product-formula results (Fig.~\ref{fig:trotter}, red symbols) on quantum device remain close to results on a classical computer, the product formula no longer provides a good approximation to the exact dynamics, as can be seen from the deviations of the open red diamonds from the black dashed lines. 
The approximation becomes worse at even higher orders (not shown), which require exponentially more gates per step.

It is worth emphasizing that product formulas indeed become more accurate at higher orders, provided that the time step size $\Delta$ is fixed. In our tests, higher order algorithms perform worse because $\Delta$ is changed, such that the total gate depth does not exceed what is viable on the quantum device. 
An analogy here is the run time on classical computers. While higher order algorithms are more accurate at a fixed resolution, they require more operations and therefore longer run time. When given a fixed run time, one is forced to use a coarser resolution, in which case higher order algorithms may perform worse than lower order algorithms.

\subsection{Optimal use of limited quantum resources}
On current quantum devices, which do not yet have operational error correction, the maximum gate depth is limited. To make the best use of the limited quantum resources, we can adjust the choice of algorithms and resolutions for a given problem. 
In our case, the pulse compression problem seeks to determine the final seed laser intensity at the end of the interactions. The final time is set, for example, by the duration of laser pulses or the time to traverse the size of the mediating plasma. 
As a test problem, we fix $\tau_f=1$ with parameter values $\rho=4$, $\theta=0$, $s_2=3$, and $s_3=3$.
We perform Hamiltonian simulation on the quantum device using product formulas to evolve quantum states, and the measured occupation probabilities are post-processed to compute the expectation values of the three waves using Eqs.~(\ref{eq:expectation}).
We measure the error of the simulation by $\epsilon=\{\frac{1}{N} \sum_{k=1}^N [n(k\Delta)-\langle n(k\Delta)\rangle]^2\}^{1/2}$, where $n$ is the exact result on a classical computer and $\langle n\rangle$ is the expectation value obtained from the quantum hardware. In other words, we define the overall error of the simulation to be the 2-norm between the exact and measured time series, normalized by the number of time steps. 
The average error per step, $\epsilon$, receives higher contributions from later steps of the simulation. In the definition of $\epsilon$, it makes no difference whether we use $n_1$, $n_2$, or $n_3$, because $s_2=n_1+n_2$ and $s_3=n_1+n_3$ are exact constants of motion. Notice that the error of $\langle n\rangle$ is different from, albeit correlated with, the errors in the unitary $\mathsfbi{U}$. The unitary error, which is also known as the process infidelity, gives a more complete characterization of the hardware performance. But the expectation-value error is of more interest to the pulse compression problem: it is the same type of error that would typically be determined using a classical algorithm, and is much easier to measure than full process tomography in experiments.

The overall error receives contributions from two fundamental sources. 
First, algorithmic errors often arise when simulating a dynamical system with knowledge of only the exact solutions of its noncommuting subsystems. Algorithmic errors are unavoidable even on classical computers. In our test problem, we assume that the separate three- and four-wave unitary can be implemented exactly, and then use product formulas to approximate the total unitary. In this case, any finite time step size introduces a discretization error $\epsilon_\Delta$, which can be reduced either by using higher order formulas at fixed $\Delta$, or by using the formula at a fixed order but with decreasing $\Delta$. 
When using the Suzuki formula, demanding errors to scale as $\Delta^{q+1}$ per step requires $M_q$ operations, which grows exponentially with $q$. 
On the other hand, to reach a target final time $\tau_f$, the number of steps $N=\tau_f/\Delta$ increases only linearly when decreasing $\Delta$. 
The total algorithmic error $\epsilon_1=O(\tau_f\Delta^q)$ can in principle be made arbitrarily small by increasing $q$ and decreasing $\Delta$. However, in practice, given a limited run time, finite algorithm precision must be chosen. 
The trade-off between using a larger $q$ versus a smaller $\Delta$ is strongly influenced by the second fundamental source of error: the hardware error. 
We measure hardware error by $\epsilon_Q$, the error per unitary operation, which is analogous to round-off errors on classical computers. On future error corrected quantum computers, it will be possible to suppress $\epsilon_Q$ to arbitrarily small values. However, on current noisy devices, with only error mitigation rather than error correction, $\epsilon_Q$ is substantial. Using randomized compilation, we transform coherent errors into random Pauli errors, which contribute to depolarizing noise together with intrinsic quantum decoherence. After error mitigation, the error for one operation becomes independent from the previous operation, so the total hardware error $\epsilon_2=O(NM_q\epsilon_Q)$ accumulates linearly with the number of operations in the worst-case scenario. 
Because the algorithmic error $\epsilon_1$ is independent of the hardware error $\epsilon_2$, the overall error is $\epsilon=\epsilon_1+\epsilon_2$. Notice that $\epsilon_1$ decreases with $N$, whereas $\epsilon_2$ increases with $N$, so there is an optimal resolution $\Delta$ at which $\epsilon$ is minimized.

\begin{figure}
  \centering
  \includegraphics[width=0.75\textwidth]{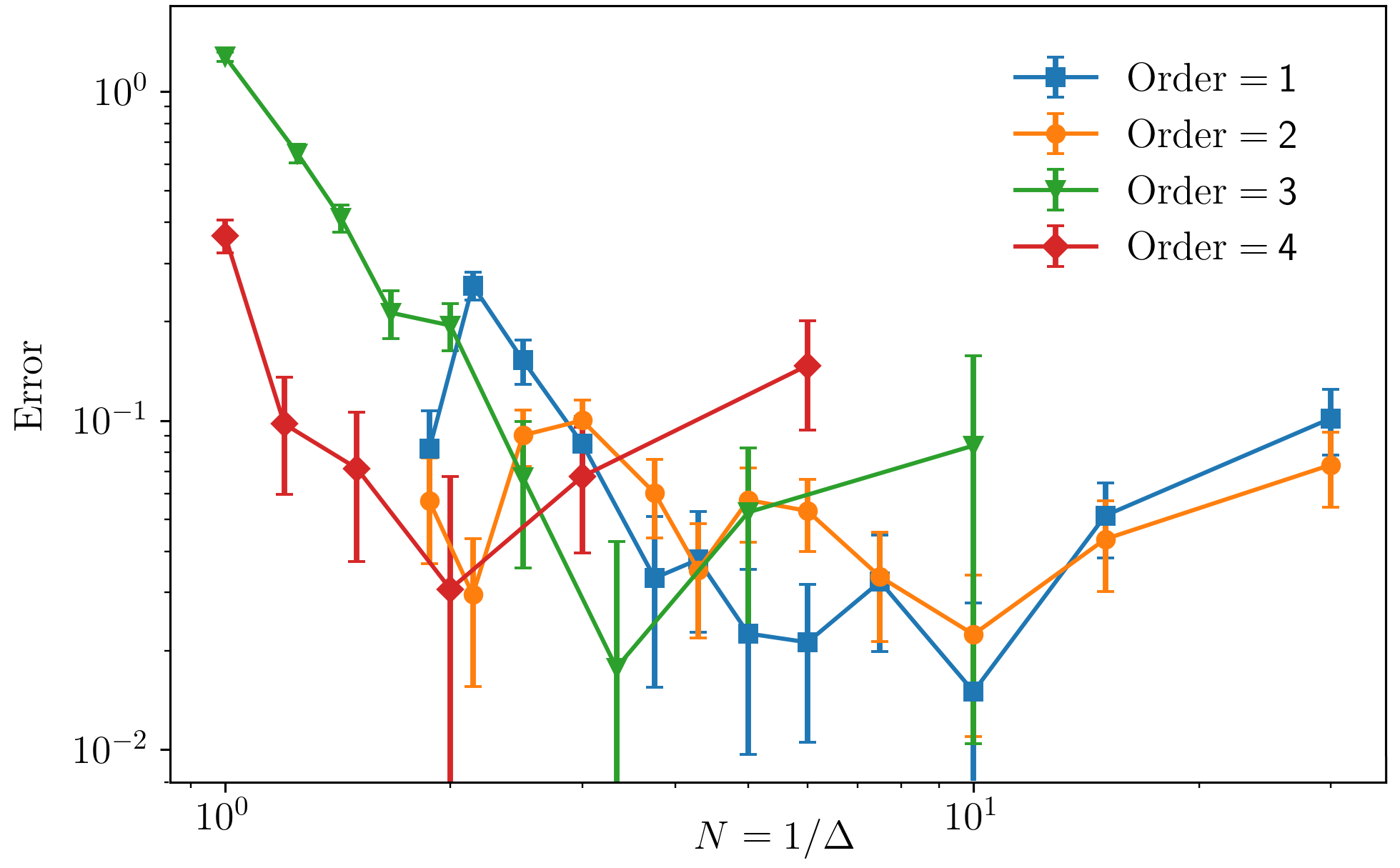}
  \caption{For a given problem, minimum error is obtained at the trade-off between algorithmic errors and hardware errors. All test problems use common parameters $\rho=4$, $\theta=0$, $s_2=3$, and $s_3=3$. The targeted final time $\tau_f=N\Delta=1$ is fixed, so a finer resolution $\Delta$ requires more steps $N$, which means algorithmic errors decrease with $N$ at the expense of accumulating more hardware errors. An optimal resolution exist, where the overall error is minimized. At higher order, the optimal $N$ shifts towards lower resolution, and the minimum error does not improve with the algorithm order. 
  }
\label{fig:tradeoff}
\end{figure}

The trade-off between hardware and algorithmic errors is demonstrated by a suite of experiments, whose results are shown in Fig.~\ref{fig:tradeoff}. We test the four product formulas using a common test problem, whose parameters are $\rho=4$, $\theta=0$, $s_2=3$, $s_3=3$, and $\tau_f=1$.
For each order of the product formula, the overall error first decreases with $N$ due to the reduction of algorithmic errors at finer resolution. However, when $N$ exceeds an optimal value $N_*$, the error starts to increase due to the accumulation of hardware errors. 
At small $N$, the resolution is too coarse to resolve the dynamics, so the errors are $O(1)$, which are comparable to the signals. 
If $N_*$ had been larger, one would expect to see that $\epsilon$ for higher order algorithms is smaller and decreases at a steeper slope. In our tests, because $N_*$ is not large enough, such a behavior is not clearly observed. 
At large $N$, where the accumulation of hardware errors dominates, $\epsilon$ increases roughly linearly with $N$ for all orders. Because higher order algorithms use more operations per step $M_q$, higher order curves reside above lower order ones in the log-log plots of $\epsilon$, except between orders $1$ and $2$. 
Notice that although $M_1=2$ and $M_2=3$, after merging adjacent unitary operations of the same type, as discussed after Eq.~(\ref{eq:U2}), the first-order sequence has $2N$ operations, while the second-order sequence has $2N+1$ operations, which is only slightly larger.
The advantage of the second-order formula is not apparent because $\epsilon$ measures the error in $n$ rather than in $\mathsfbi{U}$. On a classical computer, we observe a similar behavior: 
at small $N$, while the second-order formula has a smaller error in $\mathsfbi{U}$, it has a larger error in $n$; at large $N$, while the second-order formula has a larger error in $\mathsfbi{U}$, it has a smaller error in $n$. 
These behaviors are perhaps peculiar to our mixed three- and four-wave interaction problem. 
At intermediate $N$, higher order algorithms tend to have a smaller optimal $N_*$. This behavior can be understood from $\epsilon_1\simeq C/N^q$ and $\epsilon_2\simeq AB^qN$, where $B^q$ comes from the exponential scaling of $M_q$. The minimum of $\epsilon(N)$ is reached at $N_*=\frac{1}{B}(R q)^{1/(q+1)}$, where $R=BC/A$. 
The function $N_*(q)$ is not monotonic: it increases to a maximum before decreasing as $q^{1/q}$. 
When the hardware error is small compared to the algorithmic error, $R$ is large, in which case the maximum of $N_*(q)$ is reached at a small $q$ value. 
This means $N_*$ decreases with order, which is what we observe in our experiments.

While hardware errors are negligible in classical computers, current quantum devices have significant hardware errors. These errors make the trade-off with algorithmic errors a relevant issue. It is worth pointing out that when considering the effect of round-off errors, such a trade-off also exists for classical computers, but, due to the high precision available on today's classical computers, this is far less consequential. 
In the small hardware error limit, 
$R=BC/A\to\infty$, so $N_*$ monotonically decreases for $q\ge 1$. At a fixed $q$, $N_*$ increases with $R$, which means a smaller hardware error can support the use of a finer resolution. On a typical classical computer, $A\sim 10^{-12}$ for our test problem, so that $N_*\sim 10^4$, at which the total error $\epsilon\sim10^{-8}$ is negligible. 
In contrast, current quantum hardware has error $A\sim 10^{-2}$, which means in our test problem, the hardware can only support $N_*\sim 10^1$, for which the total error $\epsilon\sim10^{-2}$ is noticeable. In such a scenario, there is no clear benefit of using higher-order product formulas because at a given hardware error, such formulas favor the use of lower resolution. 
But when the resolution becomes too low to resolve the dynamics, the overall error becomes worse with increasing algorithm order. In our test, higher order algorithms do not perform better than a first-order algorithm at their respective optimal $N_*$. However, this may change on future quantum devices where $\epsilon_Q$ becomes even smaller. 
Similar conclusions have recently been reached for simulating the transverse-field Ising model and the XY model on noisy quantum computers \citep{avtandilyan2024optimal}.

\section{Conclusion}\label{sec:conclusion}
In summary, we develop a quantization approach for solving nonlinear wave-wave interaction problems on quantum computers. The approach becomes necessary when simulating interactions of quantum light with plasmas. The mixed three- and four-wave interaction problem serves as a nontrivial test for current quantum devices. We implement product formula algorithms using two superconducting qubits along with a suite of error mitigation techniques. The mitigated hardware error is small enough that we can begin to compare different product formulas and perform interesting simulations on quantum devices. In the future, when intrinsic hardware errors become even smaller, it may become feasible to explore more sophisticated algorithms for more nontrivial problems. 

{\bf Acknowledgements.} We thank Bhuvanesh Sundar and Joseph Andress for providing feedback to earlier versions of this paper. This work was performed under the auspices of the U.S. Department of Energy Fusion Energy Sciences by Lawrence Livermore National Laboratory under Contract DE-AC52-07NA27344 and under grant number SC-FES SCW1736. The paper is reviewed and released under LLNL-JRNL-865324.

\appendix
\section{Physically-motivated error model}\label{appA}
As discussed in Sec.~\ref{sec:mitigation}, we use twirling to convert coherent errors to depolarizing channels and rescale probabilities to compensate for exponential decays. In this appendix, we provide justifications for these mitigation steps by investigating a physically motivated error model.  
We use the technique of cycle error reconstruction to measure an empirical error model of our gate under Pauli twirling. Although our native gate is non-Clifford, we are able to propagate the 2Q inversion to the end of the circuit, meaning that the inversion appears only as a term in the SPAM coefficient of the decay. We perform the cycle error reconstruction for 1-body errors, meaning that we estimate the nine 2Q Pauli error rates $(\mathsfbi{X}\mathsfbi{X}, \mathsfbi{X}\mathsfbi{Y}, \mathsfbi{X}\mathsfbi{Z}, ...)$ and the six 1Q Pauli error rates $(\mathsfbi{X}\mathsfbi{I}, \mathsfbi{I}\mathsfbi{X}, ...)$. However, not all of these error rates are individually observable. The entangling gate induces orbitals of certain types of Pauli errors, meaning that we can only observe the product of the decay fidelities rather than individual Pauli errors.

The reconstructed Pauli error rates characterize the effective noise model under Pauli twirling with high accuracy. In order to determine a full Pauli error model, namely, a model with a probability for each type of Pauli error, a method of distributing the degenerate error rates to their components is required. This can be achieved by simply assuming, for example, that both Pauli errors contribute to the product equally ~\citep{berg_probabilistic_2022}. Alternatively, one can perform further characterization to isolate the marginal effect of error channels ~\citep{berg_techniques_2024}. The approach we take is to fit an open-system simulation to the observed Pauli error rates.

\begin{figure}
    \centering
    \includegraphics[width=0.95\linewidth]{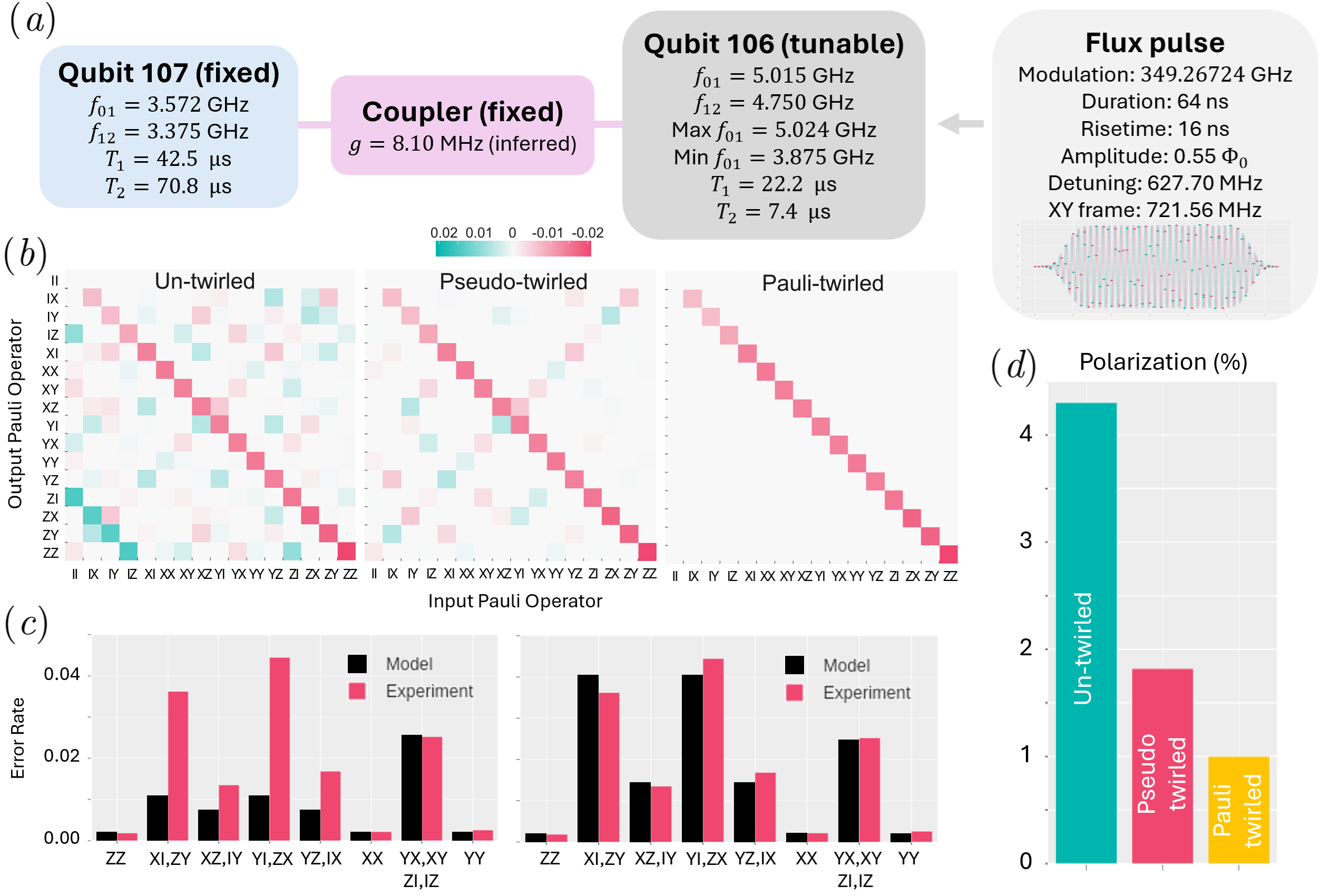}
    \caption{Modeling of our two-qubit SQISW gate reveals that errors become approximately depolarizing noise after twirling. 
    (\textit{a}) The two-qubit system we use consists of a fixed qubit and a tunable qubit, connected by a fixed coupler. We model the system using physical parameter values. 
    The SQISW gate is enacted by a flux pulse that modulates the tunable qubit. 
    (\textit{b}) We compute errors in the Pauli transfer matrix for the SQISW gate. The error is entirely incoherent with 1.3\% infidelity, but has a complex structure (left). After pseudo twirling, the infidelity does not change, but the error is simplified and symmetrized (middle). After Pauli twirling, the same infidelity manifests only as errors along the diagonal (right).
    (\textit{c}) After Pauli twirling, the observed Pauli errors for the SQISW gate (red) is not entirely explained by the model when using reported decoherence time $T_1$ and $T_2$ (left), which are measured when the gate is not in action. Nevertheless, by tuning the decoherence times, Pauli error reconstruction using our model (black) matches the observed errors (right).
    (\textit{d}) We measure polarization of errors as the diamond-norm
    distance from pure depolarizing channels of the same infidelity. Twirling reduces noise polarization, which facilitates our error mitigation.
    }
    \label{fig:ErrorSimulation}
\end{figure}

Using physical values of qubit parameters, including qubit frequencies, anharmonicities, a fixed coupling extracted from gate operation parameters, and the parameters of our calibrated pulse, we are able to represent the full two-qubit system using a Lindblad model, as detailed in Fig.~\ref{fig:ErrorSimulation}(\textit{a}). In order to ensure the model replicates the system dynamics, we simulate our calibrated SQISW gate, which is enacted by a flux pulse on the tunable qubit. We adjust the qubit-qubit coupling to match our calibration. The determined coupling is 8.1 MHz, which is consistent with the design value and the operation of the two-qubit gates.

To simulate the effect of decoherence, we adjust damping ($T_1$) and dephasing ($T_2$) terms in the Lindblad model, starting with the reported values on the device. The simulation allows us to determine an effective superoperator for the SQISW gate. The errors, namely, the distance from the superoperator of an ideal gate, are shown in the left panel of Fig.~\ref{fig:ErrorSimulation}(\textit{b}). Using the simulation, we can also apply pseudo twirling and Pauli twirling to determine the twirled superoperators, whose errors are shown in the middle and the right panels of Fig.~\ref{fig:ErrorSimulation}(\textit{b}). The Pauli-twirled superoperator can then be transformed to Pauli error rates via a Walsh-Hadamard transform, which is compared to the observed error rates. Using reported coherence times, some important aspects of the observed Pauli error rates are reproduced, as shown in the left panel of Fig.~\ref{fig:ErrorSimulation}(\textit{c}). However, the overall error rate is different and there are significant discrepancies in the error profile.

To improve our error model, we consider two additional effects. First, single-qubit gates also contribute noise to the cycle. Because single-qubit gates are randomized under twirling, we model this as a depolarizing error. Secondly, coherence decreases when qubits are under modulation \citep{didier_flux_2019}. Thus we allow the coherence times of the model to vary in order to match the observed error profile. The result is a closer match between the simulated and the observed error profiles, as shown in the right panel of Fig.~\ref{fig:ErrorSimulation}(\textit{c}). The model converges on a significantly decreased $T_1$ time for both qubits. While $T_1$ is not normally expected to decrease under modulation, other mechanisms, such as leakage or two-level system loss, might have a similar effect during Pauli error reconstruction. 

With a simulated error model in hand, we now have the ability to generate a Pauli error model that both reproduces the empirically observed Pauli error rates, and allows us to decompose the errors into individual Pauli terms. While probabilistic error amplification (PEA) and probabilistic error cancellation (PEC) strategies make use of these decompositions \citep{ferracin_efficiently_2022, berg_probabilistic_2022}, our re-scaling technique naively assumes depolarizing noise which uniformly decreases observable values. Moreover, in the characterization phase, we were able to use Pauli twirling by propagating the inversion of the benchmark circuit, but this is not possible in algorithm circuits. Thus, for our problem circuits, we use a pseudo-twirling technique which does not tailor noise as completely as Pauli twirling.

Using the physical noise model, we can now test the assumption that errors become depolarizing after twirling. As a metric of polarization, we compare the diamond-norm distance of the full superoperator, the Pauli-twirled superoperator and the pseudo-twirled superoperator to a depolarizing channel of the same infidelity. By this metric, a depolarizing channel would have a polarization of zero. As shown in Fig.~\ref{fig:ErrorSimulation}(\textit{d}), we find that the bare channel has a polarization of 4.3\%, while the Pauli-twirled channel has a polarization of 1.0\%. The pseudo-twirled channel has a polarization of 1.8\%, meaning that, as expected, pseudo-twirled channel is slightly less effective. The residual polarization implies that our rescaling technique has a systematic error, which depends on details of the problem, the observable and the noise. Nevertheless, we can conclude that for our physical noise model, Pauli twirling is effective in achieving a more depolarizing noise and that pseudo twirling retains only moderately more polarization.

In summary, we develop an approach to fitting a physical noise model to a Pauli error reconstruction which can be efficiently measured. Fitting a physical noise model to these observable error rates provides a method of inferring a full noise model based on relatively few measurements. This model can be used for error mitigation techniques, for error diagnostics, and for studying the effectiveness of different twirling groups. By introducing even more effects into the simulation and constraining their values within reasonable bounds, we expect that better agreement with observations can be achieved, thus opening the door to further error mitigation and suppression strategies. The approach outlined in this appendix is scalable to cycles of any size, provided that the noise is relatively sparse. Both the Pauli noise reconstruction and simulation techniques become intractable for subsystem sizes larger than a few qubits, emphasizing the importance of limited crosstalk to the neighbourhood of a few qubits. Finally, we note that existing PEA and PEC techniques typically rely on the circuit being executed with Pauli twirling, and work remains to generalize them to non-Clifford entangling gates, such as our SQISW gate.

\bibliographystyle{jpp}

\begin{thebibliography}{60}
	\expandafter\ifx\csname natexlab\endcsname\relax\def\natexlab#1{#1}\fi
	\def\au#1{#1} \def\ed#1{#1} \def\yr#1{#1}\def\at#1{#1}\def\jt#1{\textit{#1}}
	\def\bt#1{#1}\def\bvol#1{\textbf{#1}} \def\vol#1{#1} \def\pg#1{#1}
	\def\publ#1{#1}\def\arxiv#1{#1}\def\org#1{#1}\def\st#1{\textit{#1}}
	
	\bibitem[Abrams {\em et~al.\/}(2020)Abrams, Didier, Johnson, Silva \&
	Ryan]{abrams_implementation_2020}
	{\sc \au{Abrams, Deanna~M.}, \au{Didier, Nicolas}, \au{Johnson, Blake~R.},
		\au{Silva, Marcus P.~da} \& \au{Ryan, Colm~A.}} \yr{2020}  \at{Implementation
		of {XY} entangling gates with a single calibrated pulse}.  \jt{Nature
		Electronics}  \bvol{3}~(12),  \pg{744--750}, number: 12 Publisher: Nature
	Publishing Group.
	
	\bibitem[Aspelmeyer \& Schwab(2008)]{aspelmeyer2008focus}
	{\sc \au{Aspelmeyer, Markus} \& \au{Schwab, Keith}} \yr{2008} Focus on
	mechanical systems at the quantum limit.
	
	\bibitem[Avtandilyan \& Pogosov(2024)]{avtandilyan2024optimal}
	{\sc \au{Avtandilyan, AA} \& \au{Pogosov, WV}} \yr{2024}  \at{Optimal-order
		{T}rotter-{S}uzuki decomposition for quantum simulation on noisy quantum
		computers}.  \jt{arXiv:2405.01131}.
	
	\bibitem[Beale {\em et~al.\/}(2020)Beale, Boone, Carignan-Dugas, Chytros,
	Dahlen, Dawkins, Emerson, Ferracin, Frey, Hincks, Hufnagel, Iyer, Jain,
	Kolbush, Ospadov, Pino, Qassim, Saunders, Skanes-Norman, Stasiuk, Wallman,
	Winick \& Wright]{trueq}
	{\sc \au{Beale, Stefanie~J.}, \au{Boone, Kristine}, \au{Carignan-Dugas,
			Arnaud}, \au{Chytros, Anthony}, \au{Dahlen, Dar}, \au{Dawkins, Hillary},
		\au{Emerson, Joseph}, \au{Ferracin, Samuele}, \au{Frey, Virginia},
		\au{Hincks, Ian}, \au{Hufnagel, David}, \au{Iyer, Pavithran}, \au{Jain,
			Aditya}, \au{Kolbush, Jason}, \au{Ospadov, Egor}, \au{Pino, JosÃ©~Luis},
		\au{Qassim, Hammam}, \au{Saunders, Jordan}, \au{Skanes-Norman, Joshua},
		\au{Stasiuk, Andrew}, \au{Wallman, Joel~J.}, \au{Winick, Adam} \& \au{Wright,
			Emily}} \yr{2020} True-q.
	
	\bibitem[Berg {\em et~al.\/}(2022)Berg, Minev, Kandala \&
	Temme]{berg_probabilistic_2022}
	{\sc \au{Berg, Ewout van~den}, \au{Minev, Zlatko~K.}, \au{Kandala, Abhinav} \&
		\au{Temme, Kristan}} \yr{2022} Probabilistic error cancellation with sparse
	{Pauli}-{Lindblad} models on noisy quantum processors. arXiv:2201.09866.
	
	\bibitem[Berg \& Wocjan(2024)]{berg_techniques_2024}
	{\sc \au{Berg, Ewout van~den} \& \au{Wocjan, Pawel}} \yr{2024} Techniques for
	learning sparse {Pauli}-{Lindblad} noise models. arXiv:2311.15408.
	
	\bibitem[Berry {\em et~al.\/}(2007)Berry, Ahokas, Cleve \&
	Sanders]{berry2007efficient}
	{\sc \au{Berry, Dominic~W}, \au{Ahokas, Graeme}, \au{Cleve, Richard} \&
		\au{Sanders, Barry~C}} \yr{2007}  \at{{Efficient quantum algorithms for
			simulating sparse Hamiltonians}}.  \jt{Communications in Mathematical
		Physics}  \bvol{270},  \pg{359--371}.
	
	\bibitem[Berry {\em et~al.\/}(2014)Berry, Childs, Cleve, Kothari \&
	Somma]{berry2014exponential}
	{\sc \au{Berry, Dominic~W}, \au{Childs, Andrew~M}, \au{Cleve, Richard},
		\au{Kothari, Robin} \& \au{Somma, Rolando~D}} \yr{2014} {Exponential
		improvement in precision for simulating sparse Hamiltonians}.  \bt{In {\em
			Proceedings of the forty-sixth annual ACM symposium on Theory of
			computing\/}},  \pg{pp. 283--292}.
	
	\bibitem[Bowen {\em et~al.\/}(2018)Bowen, Badman, Hellinger \&
	Bale]{bowen2018density}
	{\sc \au{Bowen, Trevor~A}, \au{Badman, Samuel}, \au{Hellinger, Petr} \&
		\au{Bale, Stuart~D}} \yr{2018}  \at{{Density fluctuations in the solar wind
			driven by Alfv{\'e}n wave parametric decay}}.  \jt{The Astrophysical Journal
		Letters}  \bvol{854}~(2),  \pg{L33}.
	
	\bibitem[Bowen \& Milburn(2015)]{bowen2015quantum}
	{\sc \au{Bowen, Warwick~P} \& \au{Milburn, Gerard~J}} \yr{2015} {\em Quantum
		optomechanics\/}.  \publ{CRC press}.
	
	\bibitem[Breitenbach {\em et~al.\/}(1997)Breitenbach, Schiller \&
	Mlynek]{breitenbach1997measurement}
	{\sc \au{Breitenbach, Gerd}, \au{Schiller, S} \& \au{Mlynek, J}} \yr{1997}
	\at{Measurement of the quantum states of squeezed light}.  \jt{Nature}
	\bvol{387}~(6632),  \pg{471--475}.
	
	\bibitem[Brillouin(1914)]{brillouin1914light}
	{\sc \au{Brillouin, L{\'e}on}} \yr{1914}  \at{Light diffusion by a homogeneous
		transparent body}.  \jt{Comptes Rendus Hebdomadaires des Seances de
		lâ€™Academie des Sciences}  \bvol{158},  \pg{1331--1334}.
	
	\bibitem[Caldwell {\em et~al.\/}(2018)Caldwell, Didier, Ryan, Sete, Hudson,
	Karalekas, Manenti, da~Silva, Sinclair, Acala {\em
		et~al.\/}]{caldwell2018parametrically}
	{\sc \au{Caldwell, SA}, \au{Didier, N}, \au{Ryan, CA}, \au{Sete, EA},
		\au{Hudson, A}, \au{Karalekas, P}, \au{Manenti, R}, \au{da~Silva, MP},
		\au{Sinclair, R}, \au{Acala, E} \& \au{others}} \yr{2018}  \at{Parametrically
		activated entangling gates using transmon qubits}.  \jt{Physical Review
		Applied}  \bvol{10}~(3),  \pg{034050}.
	
	\bibitem[Carignan-Dugas {\em et~al.\/}(2023)Carignan-Dugas, Dahlen, Hincks,
	Ospadov, Beale, Ferracin, Skanes-Norman, Emerson \&
	Wallman]{carignan-dugas_error_2023}
	{\sc \au{Carignan-Dugas, Arnaud}, \au{Dahlen, Dar}, \au{Hincks, Ian},
		\au{Ospadov, Egor}, \au{Beale, Stefanie~J.}, \au{Ferracin, Samuele},
		\au{Skanes-Norman, Joshua}, \au{Emerson, Joseph} \& \au{Wallman, Joel~J.}}
	\yr{2023} The {Error} {Reconstruction} and {Compiled} {Calibration} of
	{Quantum} {Computing} {Cycles}. arXiv:2303.17714.
	
	\bibitem[Cho {\em et~al.\/}(2023)Cho, Beck, Castelli, Wendt, Evert, Reagor \&
	DuBois]{cho2023direct}
	{\sc \au{Cho, Yujin}, \au{Beck, Kristin~M}, \au{Castelli, Alessandro~R},
		\au{Wendt, Kyle~A}, \au{Evert, Bram}, \au{Reagor, Matthew~J} \& \au{DuBois,
			Jonathan~L}} \yr{2023}  \at{Direct pulse-level compilation of arbitrary
		quantum logic gates on superconducting qutrits}.  \jt{arXiv:2303.04261}.
	
	\bibitem[Davidson(2012)]{davidson2012methods}
	{\sc \au{Davidson, Ronald}} \yr{2012} {\em Methods in nonlinear plasma
		theory\/}.  \publ{Elsevier}.
	
	\bibitem[Didier(2019)]{didier_flux_2019}
	{\sc \au{Didier, Nicolas}} \yr{2019}  \at{Flux control of superconducting
		qubits at dynamical sweet spots}.  \jt{arXiv:1912.09416}.
	
	\bibitem[Dodin \& Startsev(2021)]{dodin2021applications}
	{\sc \au{Dodin, Ilya~Y} \& \au{Startsev, Edward~A}} \yr{2021}  \at{On
		applications of quantum computing to plasma simulations}.  \jt{Physics of
		Plasmas}  \bvol{28}~(9).
	
	\bibitem[Engel {\em et~al.\/}(2021)Engel, Smith \& Parker]{engel2021linear}
	{\sc \au{Engel, Alexander}, \au{Smith, Graeme} \& \au{Parker, Scott~E}}
	\yr{2021}  \at{Linear embedding of nonlinear dynamical systems and prospects
		for efficient quantum algorithms}.  \jt{Physics of Plasmas}  \bvol{28}~(6).
	
	\bibitem[Erhard {\em et~al.\/}(2019)Erhard, Wallman, Postler, Meth, Stricker,
	Martinez, Schindler, Monz, Emerson \& Blatt]{erhard2019characterizing}
	{\sc \au{Erhard, Alexander}, \au{Wallman, Joel~J}, \au{Postler, Lukas},
		\au{Meth, Michael}, \au{Stricker, Roman}, \au{Martinez, Esteban~A},
		\au{Schindler, Philipp}, \au{Monz, Thomas}, \au{Emerson, Joseph} \&
		\au{Blatt, Rainer}} \yr{2019}  \at{Characterizing large-scale quantum
		computers via cycle benchmarking}.  \jt{Nature communications}
	\bvol{10}~(1),  \pg{5347}.
	
	\bibitem[Evert {\em et~al.\/}(2024)Evert, Izquierdo, Sud, Hu, Grabbe, Rieffel,
	Reagor \& Wang]{evert2024syncopated}
	{\sc \au{Evert, Bram}, \au{Izquierdo, Zoe~Gonzalez}, \au{Sud, James}, \au{Hu,
			Hong-Ye}, \au{Grabbe, Shon}, \au{Rieffel, Eleanor~G}, \au{Reagor, Matthew~J}
		\& \au{Wang, Zhihui}} \yr{2024}  \at{Syncopated dynamical decoupling for
		suppressing crosstalk in quantum circuits}.  \jt{arXiv:2403.07836}.
	
	\bibitem[Ferracin {\em et~al.\/}(2022)Ferracin, Hashim, Ville, Naik,
	Carignan-Dugas, Qassim, Morvan, Santiago, Siddiqi \&
	Wallman]{ferracin_efficiently_2022}
	{\sc \au{Ferracin, Samuele}, \au{Hashim, Akel}, \au{Ville, Jean-Loup},
		\au{Naik, Ravi}, \au{Carignan-Dugas, Arnaud}, \au{Qassim, Hammam},
		\au{Morvan, Alexis}, \au{Santiago, David~I.}, \au{Siddiqi, Irfan} \&
		\au{Wallman, Joel~J.}} \yr{2022}  \at{Efficiently improving the performance
		of noisy quantum computers} Number: arXiv:2201.10672.
	
	\bibitem[Hansen {\em et~al.\/}(2017)Hansen, Nielsen, Salewski, Stejner, Stober,
	Team {\em et~al.\/}]{hansen2017parametric}
	{\sc \au{Hansen, S{\o}ren~Kjer}, \au{Nielsen, Stefan~Kragh}, \au{Salewski,
			Mirko}, \au{Stejner, M}, \au{Stober, J}, \au{Team, ASDEX~Upgrade} \&
		\au{others}} \yr{2017}  \at{Parametric decay instability near the upper
		hybrid resonance in magnetically confined fusion plasmas}.  \jt{Plasma
		Physics and Controlled Fusion}  \bvol{59}~(10),  \pg{105006}.
	
	\bibitem[Hashim {\em et~al.\/}(2020)Hashim, Naik, Morvan, Ville, Mitchell,
	Kreikebaum, Davis, Smith, Iancu, O'Brien {\em
		et~al.\/}]{hashim2020randomized}
	{\sc \au{Hashim, Akel}, \au{Naik, Ravi~K}, \au{Morvan, Alexis}, \au{Ville,
			Jean-Loup}, \au{Mitchell, Bradley}, \au{Kreikebaum, John~Mark}, \au{Davis,
			Marc}, \au{Smith, Ethan}, \au{Iancu, Costin}, \au{O'Brien, Kevin~P} \&
		\au{others}} \yr{2020}  \at{Randomized compiling for scalable quantum
		computing on a noisy superconducting quantum processor}.
	\jt{arXiv:2010.00215}.
	
	\bibitem[He {\em et~al.\/}(2015)He, Qin, Sun, Xiao, Zhang \&
	Liu]{he2015hamiltonian}
	{\sc \au{He, Yang}, \au{Qin, Hong}, \au{Sun, Yajuan}, \au{Xiao, Jianyuan},
		\au{Zhang, Ruili} \& \au{Liu, Jian}} \yr{2015}  \at{{Hamiltonian time
			integrators for Vlasov-Maxwell equations}}.  \jt{Physics of Plasmas}
	\bvol{22}~(12).
	
	\bibitem[Huang {\em et~al.\/}(2023)Huang, Wang, Wu, Ding, Ye, Kong, Zhang, Ni,
	Song, Shi, Zhao, Deng \& Chen]{huang_quantum_2023}
	{\sc \au{Huang, Cupjin}, \au{Wang, Tenghui}, \au{Wu, Feng}, \au{Ding, Dawei},
		\au{Ye, Qi}, \au{Kong, Linghang}, \au{Zhang, Fang}, \au{Ni, Xiaotong},
		\au{Song, Zhijun}, \au{Shi, Yaoyun}, \au{Zhao, Hui-Hai}, \au{Deng, Chunqing}
		\& \au{Chen, Jianxin}} \yr{2023}  \at{Quantum {Instruction} {Set} {Design}
		for {Performance}}.  \jt{Physical Review Letters}  \bvol{130}~(7),
	\pg{070601}, publisher: American Physical Society.
	
	\bibitem[Joseph(2020)]{joseph2020koopman}
	{\sc \au{Joseph, Ilon}} \yr{2020}  \at{{Koopman--von Neumann approach to
			quantum simulation of nonlinear classical dynamics}}.  \jt{Physical Review
		Research}  \bvol{2}~(4),  \pg{043102}.
	
	\bibitem[Joseph {\em et~al.\/}(2023)Joseph, Shi, Porter, Castelli, Geyko,
	Graziani, Libby \& DuBois]{joseph2023quantum}
	{\sc \au{Joseph, I}, \au{Shi, Y}, \au{Porter, MD}, \au{Castelli, AR},
		\au{Geyko, VI}, \au{Graziani, FR}, \au{Libby, SB} \& \au{DuBois, JL}}
	\yr{2023}  \at{Quantum computing for fusion energy science applications}.
	\jt{Physics of Plasmas}  \bvol{30}~(1).
	
	\bibitem[Kerr(1875)]{kerr1875xl}
	{\sc \au{Kerr, John}} \yr{1875}  \at{A new relation between electricity and
		light: Dielectrified media birefringent}.  \jt{The London, Edinburgh, and
		Dublin Philosophical Magazine and Journal of Science}  \bvol{50}~(332),
	\pg{337--348}.
	
	\bibitem[Klimov {\em et~al.\/}(2018)Klimov, Kelly, Chen, Neeley, Megrant,
	Burkett, Barends, Arya, Chiaro, Chen, Dunsworth, Fowler, Foxen, Gidney,
	Giustina, Graff, Huang, Jeffrey, Lucero, Mutus, Naaman, Neill, Quintana,
	Roushan, Sank, Vainsencher, Wenner, White, Boixo, Babbush, Smelyanskiy, Neven
	\& Martinis]{PhysRevLett.121.090502}
	{\sc \au{Klimov, P.~V.}, \au{Kelly, J.}, \au{Chen, Z.}, \au{Neeley, M.},
		\au{Megrant, A.}, \au{Burkett, B.}, \au{Barends, R.}, \au{Arya, K.},
		\au{Chiaro, B.}, \au{Chen, Yu}, \au{Dunsworth, A.}, \au{Fowler, A.},
		\au{Foxen, B.}, \au{Gidney, C.}, \au{Giustina, M.}, \au{Graff, R.},
		\au{Huang, T.}, \au{Jeffrey, E.}, \au{Lucero, Erik}, \au{Mutus, J.~Y.},
		\au{Naaman, O.}, \au{Neill, C.}, \au{Quintana, C.}, \au{Roushan, P.},
		\au{Sank, Daniel}, \au{Vainsencher, A.}, \au{Wenner, J.}, \au{White, T.~C.},
		\au{Boixo, S.}, \au{Babbush, R.}, \au{Smelyanskiy, V.~N.}, \au{Neven, H.} \&
		\au{Martinis, John~M.}} \yr{2018}  \at{Fluctuations of energy-relaxation
		times in superconducting qubits}.  \jt{Phys. Rev. Lett.}  \bvol{121},
	\pg{090502}.
	
	\bibitem[Kong {\em et~al.\/}(2018)Kong, Li, You, Xiong \& Wu]{kong2018two}
	{\sc \au{Kong, Cui}, \au{Li, Sha}, \au{You, Cai}, \au{Xiong, Hao} \& \au{Wu,
			Ying}} \yr{2018}  \at{Two-color second-order sideband generation in an
		optomechanical system with a two-level system}.  \jt{Scientific reports}
	\bvol{8}~(1),  \pg{1060}.
	
	\bibitem[Koukoutsis {\em et~al.\/}(2023)Koukoutsis, Hizanidis, Vahala, Soe,
	Vahala \& Ram]{koukoutsis2023quantum}
	{\sc \au{Koukoutsis, Efstratios}, \au{Hizanidis, Kyriakos}, \au{Vahala,
			George}, \au{Soe, Min}, \au{Vahala, Linda} \& \au{Ram, Abhay~K}} \yr{2023}
	\at{Quantum computing perspective for electromagnetic wave propagation in
		cold magnetized plasmas}.  \jt{Physics of Plasmas}  \bvol{30}~(12).
	
	\bibitem[Lake {\em et~al.\/}(2020)Lake, Mitchell, Sanders \&
	Barclay]{lake2020two}
	{\sc \au{Lake, David~P}, \au{Mitchell, Matthew}, \au{Sanders, Barry~C} \&
		\au{Barclay, Paul~E}} \yr{2020}  \at{Two-colour interferometry and switching
		through optomechanical dark mode excitation}.  \jt{Nature communications}
	\bvol{11}~(1),  \pg{2208}.
	
	\bibitem[Lin {\em et~al.\/}(2022)Lin, Lowrie, Aslangil, Suba{\c{s}}{\i} \&
	Sornborger]{lin2022koopman}
	{\sc \au{Lin, Yen~Ting}, \au{Lowrie, Robert~B}, \au{Aslangil, Denis},
		\au{Suba{\c{s}}{\i}, Yi{\u{g}}it} \& \au{Sornborger, Andrew~T}} \yr{2022}
	\at{{Koopman von Neumann mechanics and the Koopman representation: A
			perspective on solving nonlinear dynamical systems with quantum computers}}.
	\jt{arXiv:2202.02188}.
	
	\bibitem[Liu {\em et~al.\/}(2021)Liu, Kolden, Krovi, Loureiro, Trivisa \&
	Childs]{liu2021efficient}
	{\sc \au{Liu, Jin-Peng}, \au{Kolden, Herman~{\O}ie}, \au{Krovi, Hari~K},
		\au{Loureiro, Nuno~F}, \au{Trivisa, Konstantina} \& \au{Childs, Andrew~M}}
	\yr{2021}  \at{Efficient quantum algorithm for dissipative nonlinear
		differential equations}.  \jt{Proceedings of the National Academy of
		Sciences}  \bvol{118}~(35),  \pg{e2026805118}.
	
	\bibitem[Loudon(2000)]{loudon2000quantum}
	{\sc \au{Loudon, Rodney}} \yr{2000} {\em The quantum theory of light\/}.
	\publ{OUP Oxford}.
	
	\bibitem[Low \& Chuang(2017)]{low2017optimal}
	{\sc \au{Low, Guang~Hao} \& \au{Chuang, Isaac~L}} \yr{2017}  \at{{Optimal
			Hamiltonian simulation by quantum signal processing}}.  \jt{Physical review
		letters}  \bvol{118}~(1),  \pg{010501}.
	
	\bibitem[Malkin {\em et~al.\/}(1999)Malkin, Shvets \& Fisch]{malkin1999fast}
	{\sc \au{Malkin, VM}, \au{Shvets, G} \& \au{Fisch, NJ}} \yr{1999}  \at{Fast
		compression of laser beams to highly overcritical powers}.  \jt{Physical
		review letters}  \bvol{82}~(22),  \pg{4448}.
	
	\bibitem[May \& Qin(2023)]{May23}
	{\sc \au{May, Michael~Q.} \& \au{Qin, Hong}} \yr{2023}  \at{Quantum three-wave
		instability}.  \jt{Phys. Rev. A}  \bvol{107},  \pg{062204}.
	
	\bibitem[Michel(2023)]{michel2023introduction}
	{\sc \au{Michel, Pierre}} \yr{2023} {\em Introduction to laser-plasma
		interactions\/}.  \publ{Springer Nature}.
	
	\bibitem[Morrison(2017)]{morrison2017structure}
	{\sc \au{Morrison, Philip~J}} \yr{2017}  \at{Structure and structure-preserving
		algorithms for plasma physics}.  \jt{Physics of Plasmas}  \bvol{24}~(5).
	
	\bibitem[M{\"u}ller {\em et~al.\/}(2019)M{\"u}ller, Cole \&
	Lisenfeld]{muller2019towards}
	{\sc \au{M{\"u}ller, Clemens}, \au{Cole, Jared~H} \& \au{Lisenfeld,
			J{\"u}rgen}} \yr{2019}  \at{Towards understanding two-level-systems in
		amorphous solids: insights from quantum circuits}.  \jt{Reports on Progress
		in Physics}  \bvol{82}~(12),  \pg{124501}.
	
	\bibitem[Nachman {\em et~al.\/}(2020)Nachman, Urbanek, de~Jong \&
	Bauer]{nachman2020unfolding}
	{\sc \au{Nachman, Benjamin}, \au{Urbanek, Miroslav}, \au{de~Jong, Wibe~A} \&
		\au{Bauer, Christian~W}} \yr{2020}  \at{Unfolding quantum computer readout
		noise}.  \jt{npj Quantum Information}  \bvol{6}~(1),  \pg{84}.
	
	\bibitem[Nazarenko(2011)]{nazarenko2011waveturbulence}
	{\sc \au{Nazarenko, Sergey}} \yr{2011} {\em Wave Turbulence\/}.
	
	\bibitem[Peterson {\em et~al.\/}(2020)Peterson, Crooks \&
	Smith]{peterson_fixed-depth_2020}
	{\sc \au{Peterson, Eric~C.}, \au{Crooks, Gavin~E.} \& \au{Smith, Robert~S.}}
	\yr{2020}  \at{Fixed-{Depth} {Two}-{Qubit} {Circuits} and the {Monodromy}
		{Polytope}}.  \jt{Quantum}  \bvol{4},  \pg{247}, arXiv:1904.10541.
	
	\bibitem[Raman \& Krishnan(1928)]{raman1928negative}
	{\sc \au{Raman, CV} \& \au{Krishnan, KS}} \yr{1928}  \at{The negative
		absorption of radiation}.  \jt{Nature}  \bvol{122}~(3062),  \pg{12--13}.
	
	\bibitem[Shang(2023)]{shang2023coupling}
	{\sc \au{Shang, Cheng}} \yr{2023}  \at{Coupling enhancement and symmetrization
		of single-photon optomechanics in open quantum systems}.
	\jt{arXiv:2302.04897}.
	
	\bibitem[Shi {\em et~al.\/}(2021)Shi, Castelli, Wu, Joseph, Geyko, Graziani,
	Libby, Parker, Rosen, Martinez {\em et~al.\/}]{shi2021simulating}
	{\sc \au{Shi, Yuan}, \au{Castelli, Alessandro~R}, \au{Wu, Xian}, \au{Joseph,
			Ilon}, \au{Geyko, Vasily}, \au{Graziani, Frank~R}, \au{Libby, Stephen~B},
		\au{Parker, Jeffrey~B}, \au{Rosen, Yaniv~J}, \au{Martinez, Luis~A} \&
		\au{others}} \yr{2021}  \at{Simulating non-native cubic interactions on noisy
		quantum machines}.  \jt{Physical Review A}  \bvol{103}~(6),  \pg{062608}.
	
	\bibitem[Shi {\em et~al.\/}(2017)Shi, Qin \& Fisch]{shi2017three}
	{\sc \au{Shi, Yuan}, \au{Qin, Hong} \& \au{Fisch, Nathaniel~J}} \yr{2017}
	\at{Three-wave scattering in magnetized plasmas: From cold fluid to quantized
		{L}agrangian}.  \jt{Physical Review E}  \bvol{96}~(2),  \pg{023204}.
	
	\bibitem[Smith {\em et~al.\/}(2020)Smith, Peterson, Skilbeck \&
	Davis]{smith2020open}
	{\sc \au{Smith, Robert~S}, \au{Peterson, Eric~C}, \au{Skilbeck, Mark~G} \&
		\au{Davis, Erik~J}} \yr{2020}  \at{An open-source, industrial-strength
		optimizing compiler for quantum programs}.  \jt{Quantum Science and
		Technology}  \bvol{5}~(4),  \pg{044001}.
	
	\bibitem[Suzuki(1976)]{suzuki1976generalized}
	{\sc \au{Suzuki, Masuo}} \yr{1976}  \at{Generalized {T}rotter's formula and
		systematic approximants of exponential operators and inner derivations with
		applications to many-body problems}.  \jt{Communications in Mathematical
		Physics}  \bvol{51}~(2),  \pg{183--190}.
	
	\bibitem[Suzuki(1990)]{suzuki1990fractal}
	{\sc \au{Suzuki, Masuo}} \yr{1990}  \at{Fractal decomposition of exponential
		operators with applications to many-body theories and monte carlo
		simulations}.  \jt{Physics Letters A}  \bvol{146}~(6),  \pg{319--323}.
	
	\bibitem[Tripathi {\em et~al.\/}(2022)Tripathi, Chen, Khezri, Yip,
	Levenson-Falk \& Lidar]{tripathi2021suppression}
	{\sc \au{Tripathi, Vinay}, \au{Chen, Huo}, \au{Khezri, Mostafa}, \au{Yip,
			Ka-Wa}, \au{Levenson-Falk, E.~M.} \& \au{Lidar, Daniel~A.}} \yr{2022}
	\at{Suppression of crosstalk in superconducting qubits using dynamical
		decoupling}.  \jt{Physical Review Applied}  \bvol{18}~(2),  \pg{024068}.
	
	\bibitem[Trotter(1959)]{trotter1959product}
	{\sc \au{Trotter, Hale~F}} \yr{1959}  \at{On the product of semi-groups of
		operators}.  \jt{Proceedings of the American Mathematical Society}
	\bvol{10}~(4),  \pg{545--551}.
	
	\bibitem[Veps{\"a}l{\"a}inen {\em et~al.\/}(2020)Veps{\"a}l{\"a}inen, Karamlou,
	Orrell, Dogra, Loer, Vasconcelos, Kim, Melville, Niedzielski, Yoder {\em
		et~al.\/}]{vepsalainen2020impact}
	{\sc \au{Veps{\"a}l{\"a}inen, Antti~P}, \au{Karamlou, Amir~H}, \au{Orrell,
			John~L}, \au{Dogra, Akshunna~S}, \au{Loer, Ben}, \au{Vasconcelos, Francisca},
		\au{Kim, David~K}, \au{Melville, Alexander~J}, \au{Niedzielski, Bethany~M},
		\au{Yoder, Jonilyn~L} \& \au{others}} \yr{2020}  \at{Impact of ionizing
		radiation on superconducting qubit coherence}.  \jt{Nature}
	\bvol{584}~(7822),  \pg{551--556}.
	
	\bibitem[Ville {\em et~al.\/}(2022)Ville, Morvan, Hashim, Naik, Lu, Mitchell,
	Kreikebaum, O'Brien, Wallman, Hincks, Emerson, Smith, Younis, Iancu, Santiago
	\& Siddiqi]{Ville22}
	{\sc \au{Ville, Jean-Loup}, \au{Morvan, Alexis}, \au{Hashim, Akel}, \au{Naik,
			Ravi~K.}, \au{Lu, Marie}, \au{Mitchell, Bradley}, \au{Kreikebaum, John-Mark},
		\au{O'Brien, Kevin~P.}, \au{Wallman, Joel~J.}, \au{Hincks, Ian}, \au{Emerson,
			Joseph}, \au{Smith, Ethan}, \au{Younis, Ed}, \au{Iancu, Costin},
		\au{Santiago, David~I.} \& \au{Siddiqi, Irfan}} \yr{2022}  \at{Leveraging
		randomized compiling for the quantum imaginary-time-evolution algorithm}.
	\jt{Phys. Rev. Res.}  \bvol{4},  \pg{033140}.
	
	\bibitem[Wallman \& Emerson(2016)]{Wallman16}
	{\sc \au{Wallman, Joel~J.} \& \au{Emerson, Joseph}} \yr{2016}  \at{Noise
		tailoring for scalable quantum computation via randomized compiling}.
	\jt{Phys. Rev. A}  \bvol{94},  \pg{052325}.
	
	\bibitem[Welch {\em et~al.\/}(2014)Welch, Greenbaum, Mostame \&
	Aspuru-Guzik]{Welch_2014}
	{\sc \au{Welch, Jonathan}, \au{Greenbaum, Daniel}, \au{Mostame, Sarah} \&
		\au{Aspuru-Guzik, Alan}} \yr{2014}  \at{Efficient quantum circuits for
		diagonal unitaries without ancillas}.  \jt{New Journal of Physics}
	\bvol{16}~(3),  \pg{033040}.
	
	\bibitem[Zhakarov {\em et~al.\/}(1992)Zhakarov, L'vov \&
	Falkovich]{zakharov1992kolmogorov}
	{\sc \au{Zhakarov, Vladimir~E}, \au{L'vov, Victor~S.} \& \au{Falkovich,
			Gregory}} \yr{1992} {\em Kolmogorov Spectra of Turbulence I, Wave
		Turbulence\/}.  \publ{Springer-Verlag}.
	
	\bibitem[Zhou {\em et~al.\/}(2023)Zhou, Sitler, Oda, Schultz \&
	Quiroz]{Zeyuan:22}
	{\sc \au{Zhou, Zeyuan}, \au{Sitler, Ryan}, \au{Oda, Yasuo}, \au{Schultz, Kevin}
		\& \au{Quiroz, Gregory}} \yr{2023}  \at{Quantum crosstalk robust quantum
		control}.  \jt{Physical Review Letters}  \bvol{131}~(21),  \pg{210802}.
	
\end{thebibliography}

\end{document}